\documentclass[aps,prd,superscriptaddress,preprintnumbers]{revtex4}
\usepackage{graphicx}

\newcommand*{\ie}{\textit{i.e.},\ }

\newcommand*{\gev}{\,\textrm{GeV}}
\newcommand*{\gevc}{\,\textrm{GeV/c}}
\newcommand*{\mevc}{\,\textrm{MeV/c}}
\newcommand*{\tev}{\,\textrm{TeV}}

\newcommand*{\pt}{$p_T$}

\newcommand*{\TR}{``transverse"}

\newcommand*{\ptone}{$P_T({\rm jet}\#1)$}

\newcommand*{\Jone}{jet\#$1$}

\newcommand*{\CJone}{chgjet\#$1$}
\newcommand*{\ptcj}{$P_T$(chgjet\#$1)$}
\newcommand*{\pthard}{$p_T({\rm hard})$}

\newcommand*{\ptlcut}{$p_T\!>\!0.5\,{\rm GeV/c}$}

\newcommand*{\etacut}{$|\eta|\!<\!2$}
\newcommand*{\etacdf}{$|\eta|\!<\!1$}

\newcommand*{\etaphi}{$\eta$-$\phi$}

\begin{document}
\title{Min-Bias and the Underlying Event at the LHC}
\author{Rick Field}
\thanks{Lectures presented at the 51$^{st}$ Cracow School of Theoretical Physics:
\textit{The Soft Side of the LHC}.}

\affiliation{Department of Physics, University of Florida, Gainesville, Florida, 32611, USA}
\date{October 1, 2011}

\begin{abstract}
In a very short time the experiments at the LHC have collected a large amount of data that 
can be used to study minimum bias (MB) collisions and the underlying event (UE) in great detail.  
The CDF PYTHIA $6.2$ Tune DW predictions for the LHC UE data at $900\gev$ and $7\tev$ are examined 
in detail.  The behavior of the UE at the LHC is roughly what we expected. The LHC PYTHIA $6.4$ 
Tune Z1 does an excellent job describing the LHC UE data.  The modeling of MB (\ie the overall 
inelastic cross section) is more complicated because one must include a model of diffraction.  
The ability of PYTHIA Tune DW and Tune Z1 to simultaneously describe both the UE in a hard scattering 
process and MB collisions are studied.  No model describes perfectly all the features of MB 
collisions at the LHC.
\end{abstract}
\maketitle
  
\section{Introduction}
The total proton-proton cross section is the sum of the elastic and inelastic components, $\sigma_{tot}=\sigma_{\rm EL}+\sigma_{\rm INEL}$.  
Three distinct processes contribute to the inelastic cross section; single diffraction, double-diffraction, and everything else, 
which is referred to as ``non-diffractive" (ND) component.  For elastic scattering neither of the beam particles breaks apart 
(\ie color singlet exchange).  For single and double diffraction one or both of the beam particles are excited into a high 
mass color singlet state (\ie N* states) which then decays.  Single and double diffraction also corresponds to color 
singlet exchange between the beam hadrons.  When color is exchanged the outgoing remnants are no longer color singlets 
and one has a separation of color resulting in a multitude of quark-antiquark pairs being pulled out of the vacuum.  
The non-diffractive component, $\sigma_{\rm ND}$, involves color exchange and the separation of color. However, 
the non-diffractive collisions have both a soft and hard component.   Most of the time the color exchange between 
partons in the beam hadrons occurs through a soft interaction (\ie no high transverse momentum) and the two beam hadrons ``ooze" 
through each other producing lots of soft particles with a uniform distribution in rapidity and many particles flying down 
the beam pipe.  Occasionally, there is a hard scattering among the constituent partons producing outgoing particles and ``jets" 
with high transverse momentum.

\begin{figure}[htbp]
\begin{center}
\includegraphics[scale=0.8]{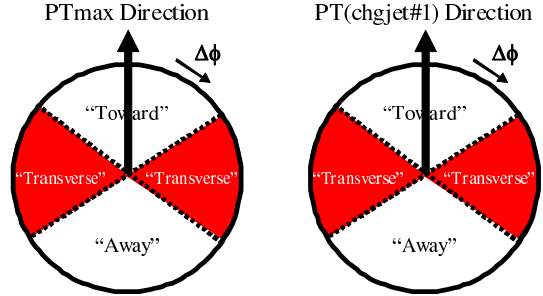}
\caption{\footnotesize
Illustration of correlations in azimuthal angle $\Delta\phi$ relative to (\textit{left}) the direction of the leading 
charged particle, PTmax, or to (\textit{right})  the leading charged particle jet, \CJone. The relative 
angle $\Delta\phi=\phi-\phi_1$, where $\phi_1$ is the azimuthal angle of PTmax (or \CJone) and $\phi$ is the 
azimuthal angle of a charged particle.  There are two \TR\ regions $60^\circ<\Delta\phi< 120^\circ$, $|\eta|<\eta_{cut}$ 
and $60^\circ<-\Delta\phi< 120^\circ$, $|\eta|<\eta_{cut}$.  The overall \TR\ region of \etaphi\ space is defined 
by $60^\circ<|\Delta\phi|< 120^\circ$, $|\eta|<\eta_{cut}$. The \TR\ charged particle density is the number of 
charged particles in the \TR\ region divided by the area in \etaphi\ space.  Similarly, the \TR\ charged PTsum 
density is the scalar PTsum of charged particles in the \TR\ region divided by the area in \etaphi\ space.
}
\end{center}
\label{Zakopane_fig1}
\end{figure}
\begin{figure}[htbp]
\begin{center}
\includegraphics[scale=0.75]{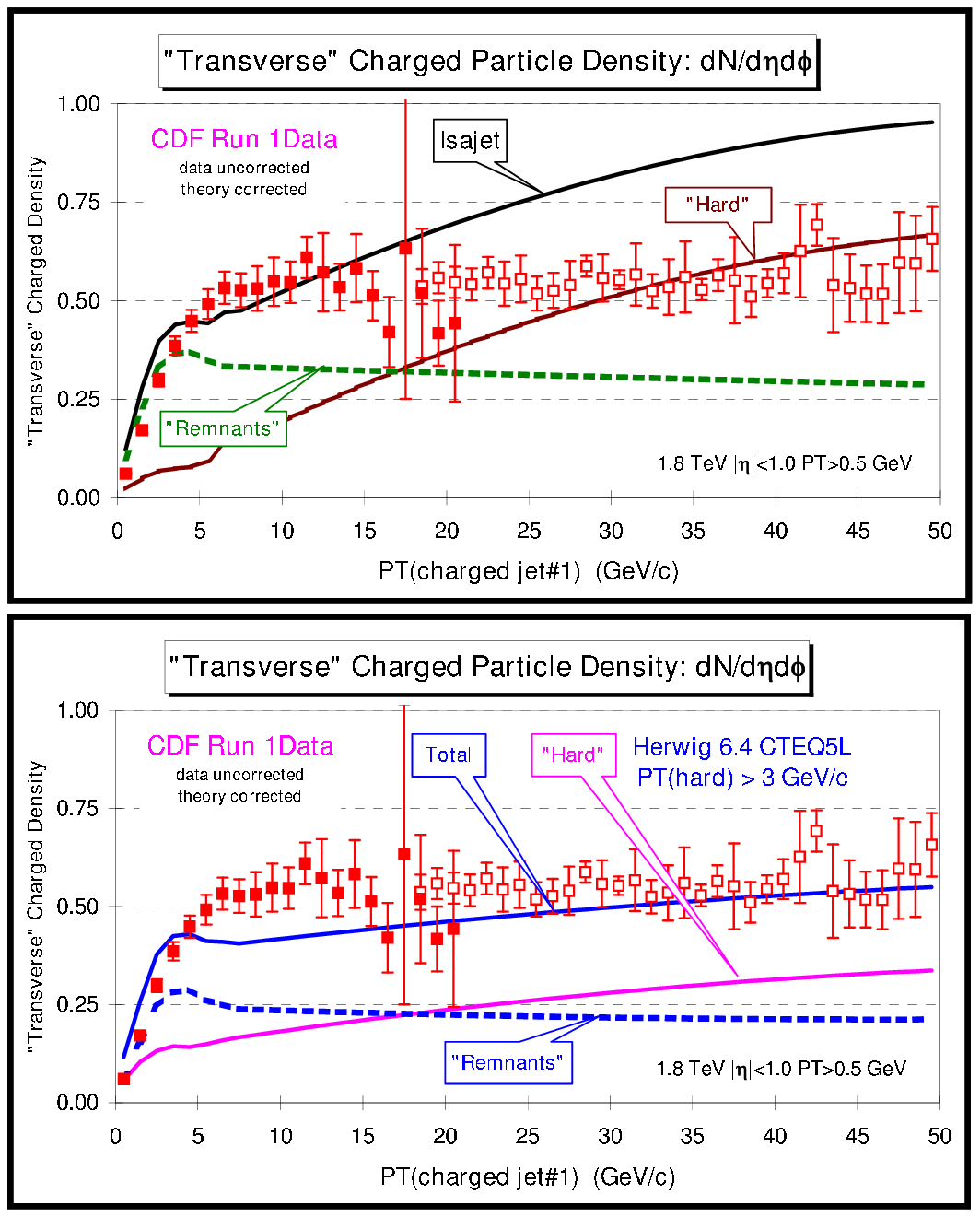}
\caption{\footnotesize
CDF Run 1 data from Ref. \cite{cdfue1} at $1.8\tev$ on the density of charged particles (\ptlcut, \etacdf) in the \TR\ region 
as defined by the leading charged particle jet, \CJone, as a function of \ptcj.  The data are compared with 
ISAJET $7.32$ without MPI (\textit{top}) and HERWIG $6.4$ without MPI (\textit{bottom}) using the ISAJET and 
HERWIG default parameters with \pthard$>3\gevc$.  The Monte-Carlo predictions are divided into two categories: 
charged particles that arise from the break-up of the beam and target (beam-beam remnants); and charged particles 
that arise from the outgoing jets plus initial and final-state radiation (hard scattering component).
}
\end{center}
\label{Zakopane_fig2}
\end{figure}

\begin{figure}[htbp]
\begin{center}
\includegraphics[scale=0.75]{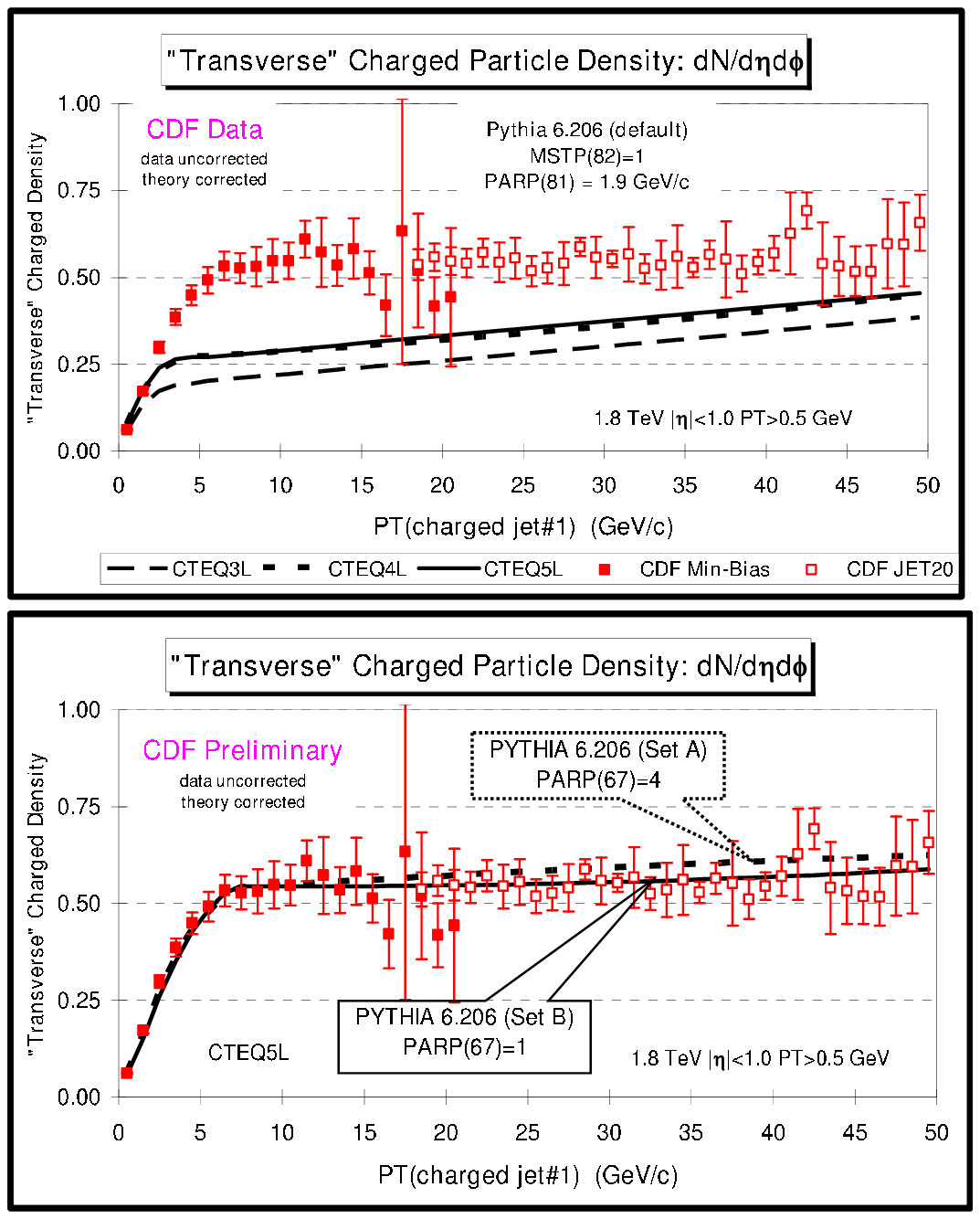}
\caption{\footnotesize
CDF Run 1 data from Ref. \cite{cdfue1} at $1.8\tev$ on the density of charged particles (\ptlcut, \etacdf) in the \TR\ region 
as defined by the leading charged particle jet, \CJone, as a function of \ptcj.  The data are compared with PYTHIA $6.206$ 
with MPI (\textit{top}) using the PYTHIA default parameters with \pthard$\ge0$ with the CTEQ3L, CTEQ4L, and CTEQ5L 
parton distribution functions.  (\textit{bottom}) Two CDF PYTHIA $6.2$ tunes, Tune A and Tune B.  Tune A was adjusted 
to fit the CDF Run 1 data with PARP($67$) $=4.0$ and Tune B was adjusted to fit the same data but with PARP($67$) $=1.0$.
}
\end{center}
\label{Zakopane_fig3}
\end{figure}

\begin{figure}[htbp]
\begin{center}
\includegraphics[scale=0.8]{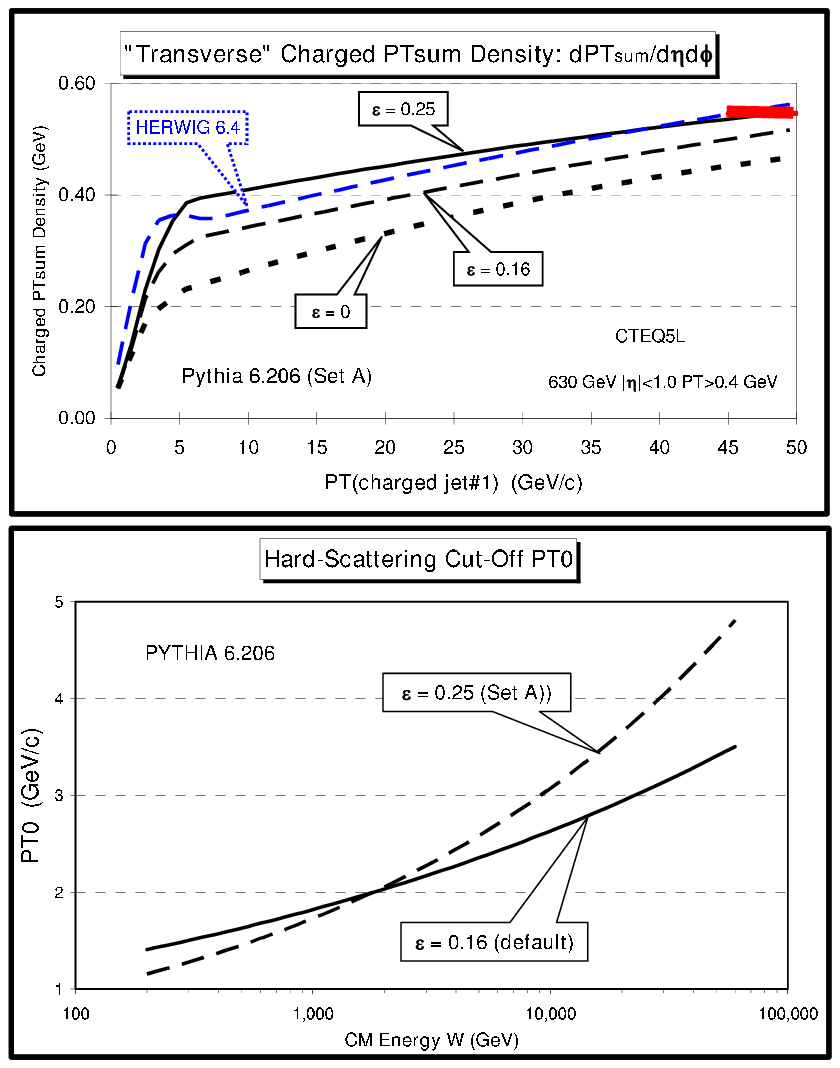}
\caption{\footnotesize
(\textit{top}) PYTHIA Tune A predictions at $630\gev$ for the charged PTsum density ($p_T\!>\!0.4\,{\rm GeV/c}$, \etacdf) in 
the \TR\ region as defined by the leading charged particle jet,  \CJone, as a function of \ptcj\ with 
$\epsilon=$ PARP($90$) $=0.0$, $0.16$ (default), and $0.25$. The CDF Run 1 data at $630\gev$ from Ref. \cite{cdf630} 
indicated a value of the PTsum density of around $0.54\gevc$ at \ptcj $\approx50\gevc$ (red line) which favors 
the PARP($90$) $=0.25$ curve. (\textit{bottom}) Shows the $2$-to-$2$ hard scattering cut-off, $p_{T0}$, versus 
center-of-mass energy from PYTHIA Tune A with the default value PARP($90$) $=0.16$ and the Tune A value of PARP($90$) $=0.25$.
}
\end{center}
\label{Zakopane_fig4}
\end{figure}

\begin{figure}[htbp]
\begin{center}
\includegraphics[scale=0.75]{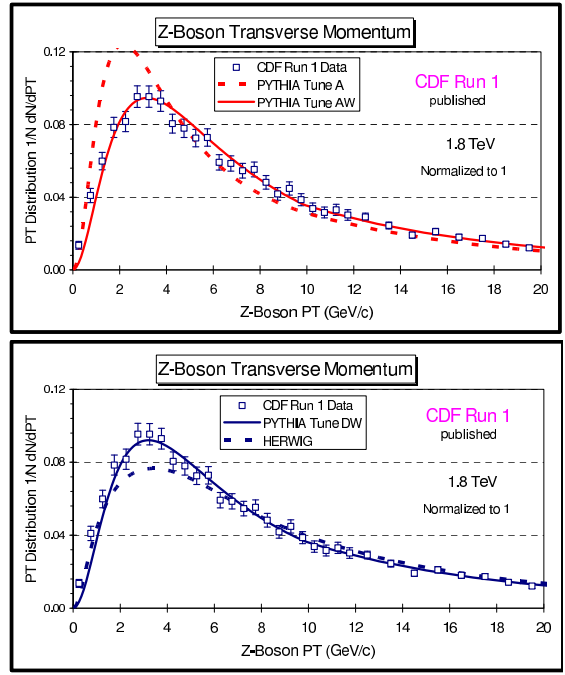}
\caption{\footnotesize
The CDF Run 1 data from Ref. \cite{cdfzpt} on the Z-boson \pt\ distribution ($<\!p_T(Z)\!>\approx11.5\gevc$) compared with 
(\textit{top}) PYTHIA Tune A ($<\!p_T(Z)\!>=9.7\gevc$) and PYTHIA Tune AW ($<\!p_T(Z)\!>=11.7\gevc$) and compared with 
(\textit{bottom}) PYTHIA Tune DW and HERWIG $6.4$ (without MPI).
}
\end{center}
\label{Zakopane_fig5}
\end{figure}

\begin{figure}[htbp]
\begin{center}
\includegraphics[scale=0.75]{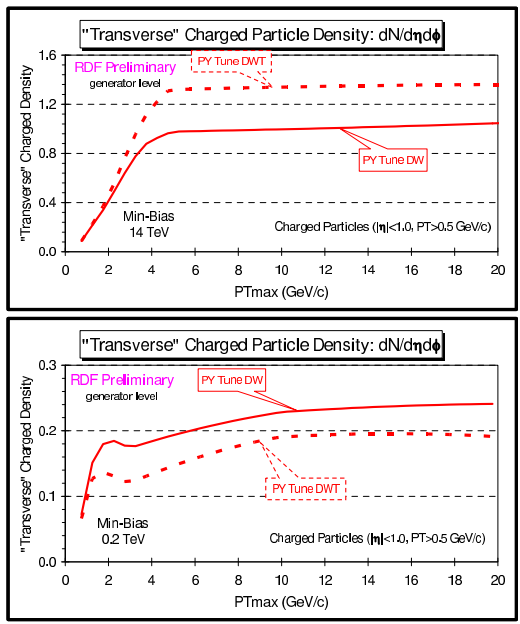}
\caption{\footnotesize
Shows the charged particle density in the \TR\ region as defined by the leading charged particle, PTmax, as function of 
PTmax for charged particles with \ptlcut\ and  \etacdf\ at $0.2\tev$ and $14\tev$ from PYTHIA Tune DW and Tune DWT at 
the particle level.  The STAR data from RHIC \cite{star} favor the energy dependence of Tune DW.
}
\end{center}
\label{Zakopane_fig6}
\end{figure}

\begin{figure}[htbp]
\begin{center}
\includegraphics[scale=0.75]{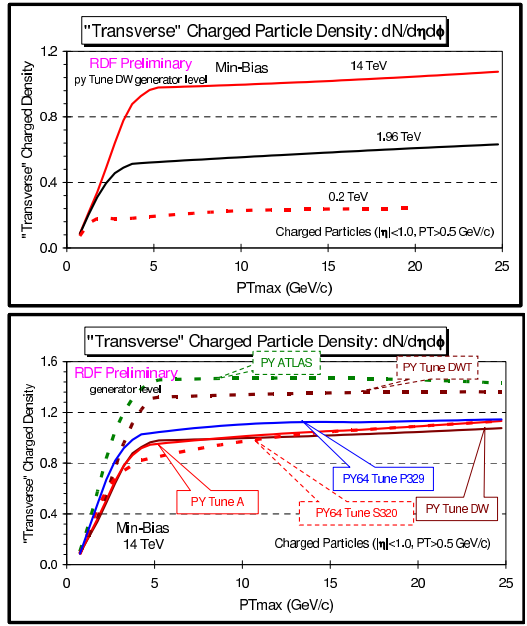}
\caption{\footnotesize
(\textit{top}) Shows the predictions at the particle level of PYTHIA Tune DW for the charged particle density in the \TR\ region 
as defined by the leading charged particle, PTmax, as a function of PTmax for charged particles (\ptlcut, \etacdf) at $0.2\tev$, 
$1.96\tev$ and $14\tev$.  (\textit{bottom}) Shows the extrapolations of PYTHIA Tune A, Tune DW, Tune DWT, Tune S320, 
Tune P329, and pyATLAS to the LHC at $14\tev$.
}
\end{center}
\label{Zakopane_fig7}
\end{figure}
\begin{figure}[htbp]
\begin{center}
\includegraphics[scale=0.6]{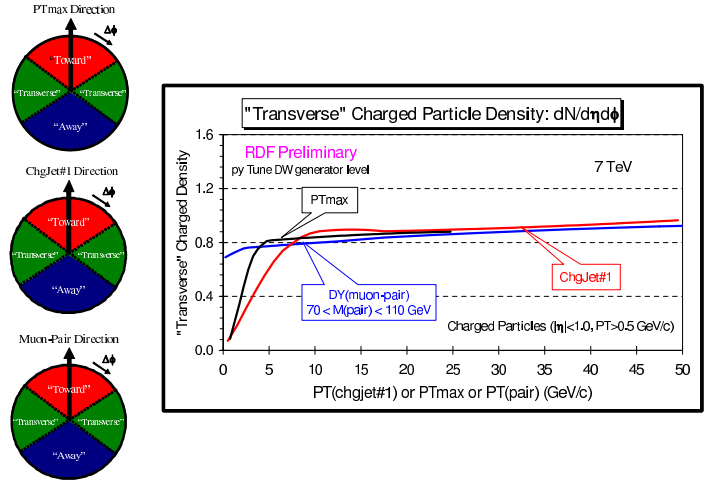}
\caption{\footnotesize
Shows the charged particle density in the \TR\ region for charged particles (\ptlcut, \etacdf) at 7 TeV as defined 
by PTmax, \ptcj, and $P_T$(muon-pair) predicted from PYTHIA Tune DW at the particle level.  For muon-pair production 
the two muons are excluded from the charged particle density.  Charged particle jets are constructed using the 
Anti-KT algorithm with $d=0.5$.
}
\end{center}
\label{Zakopane_fig8}
\end{figure}

\begin{figure}[htbp]
\begin{center}
\includegraphics[scale=0.6]{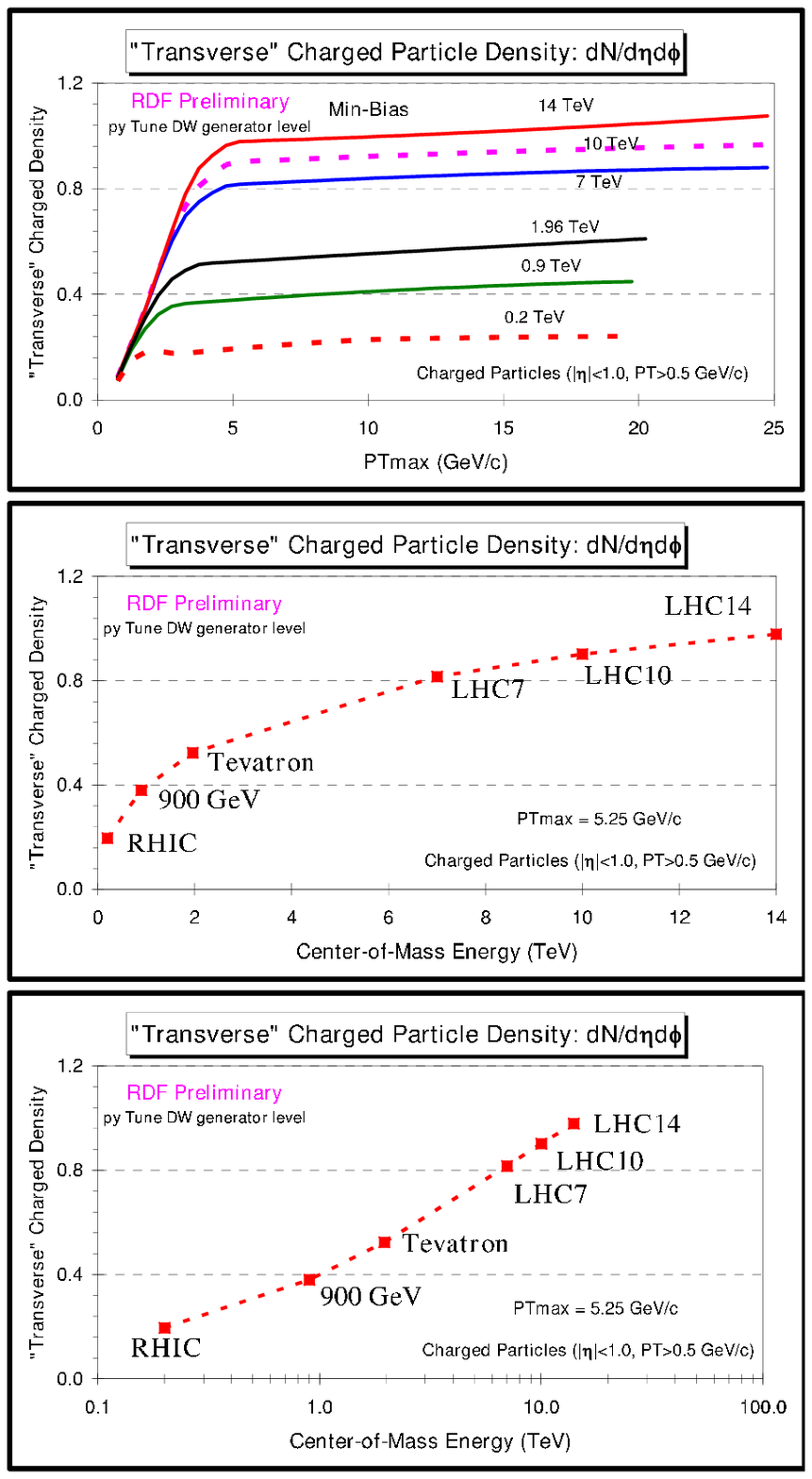}
\caption{\footnotesize
(\textit{top}) Shows the charged particle density in the \TR\ region as defined by the leading charged particle, PTmax, 
a function of PTmax for charged particles (\ptlcut, \etacdf) at $0.2\tev$, $0.9\tev$, $1.96\tev$, $7\tev$, $10\tev$, 
and $14\tev$ as predicted by PYTHIA Tune DW at the particle level. Also, shows the charged particle density in 
the \TR\ region as defined by the leading charged particle, PTmax, at PTmax $=5.25\gevc$ for charged particles (\ptlcut, \etacdf) 
as a function of the center-of-mass energy on a linear plot (\textit{middle}) and a logarithmic plot (\textit{bottom}).
}
\end{center}
\label{Zakopane_fig9}
\end{figure}

Min-bias (MB) is a generic term which refers to events that are selected with a ``loose" trigger that accepts a large fraction 
of the overall inelastic cross section.  All triggers produce some bias and the term ``min-bias" is meaningless until 
one specifies the precise trigger used to collect the data.  The underlying event (UE) consists of particles that accompany a hard scattering 
such as the beam-beam remnants (BBR) and the particles that arise from multiple parton interactions (MPI) .  The UE is 
an unavoidable background to hard-scattering collider events.  MB and UE are not the same object!  The majority of MB 
collisions are soft while the UE is studied in events in which a hard-scattering has occurred. One uses the``jet" structure of 
the hard hadron-hadron collision to experimentally study the UE \cite{cdfue1}.  As shown in Fig.~1, on an event-by-event bases, the leading 
charged particle, PTmax, or the leading charged particle jet, \CJone, can be used to isolate regions of \etaphi\ space that 
are very sensitive to the modeling of the UE.  The pseudo-rapidity is defined by $\eta=-\log(\tan(\theta_{cm}/2))$, where $\theta_{cm}$ is the 
center-of-mass polar scattering angle and $\phi$ is the azimuthal angle of outgoing charged particles.  In particular, 
the \TR\ region defined by $60^\circ<|\Delta\phi|< 120^\circ$, where $\Delta\phi=\phi-\phi_1$, where $\phi_1$ is the azimuthal 
angle of PTmax (or \CJone) and $\phi$ is the azimuthal angle of a charged particle is roughly perpendicular to the 
plane of the hard $2$-to-$2$ parton-parton scattering and is therefore very sensitive to the UE.

QCD Monte-Carlo generators such as PYTHIA \cite{pythia6} have parameters which may be adjusted to control the behavior of their event modeling.  
A specified set of these parameters that has been adjusted to better fit some aspects of the data is referred to as a tune \cite{skands1}.  
In Section 2, I will review briefly the CDF PYTHIA $6.2$ tunes.  PYTHIA Tune DW does a very nice job in describing the CDF Run 2 
underlying event data.  In Section 3, we will take a close look at how well PYTHIA Tune DW did at predicting the behavior of 
the UE at $900\gev$ and $7\tev$ at the LHC.  We will see that Tune DW does a fairly good job in describing the LHC UE data.  
However, Tune DW does not reproduce perfectly all the features of the data and after seeing the data one can construct improved 
LHC UE tunes.  The first ATLAS LHC tune was Tune AMBT1 \cite{ambt1} and CMS has Tune Z1 and Tune Z2 which I will discuss in Section 4.   
MB and the UE are not the same object, however, in PYTHIA the modeling of the UE in a hard scattering process and the modeling 
of the inelastic non-diffractive cross section are related.  In Section 5, I will examine how well the PYTHIA tunes fit the 
LHC MB data.  The summary and conclusions are in Section 6.

\section{Studying the UE at CDF}

Figure~2 shows some of the first comparisons to come from the CDF UE studies \cite{cdfue1}.  The CDF Run 1 data at $1.8\tev$ on the density of charged 
particles in the \TR\ region as defined by the leading charged particle jet, \CJone, are compared with ISAJET $7.32$ \cite{isajet} (without MPI) 
and HERWIG $6.4$ \cite{herwig} (without MPI) using the ISAJET and HERWIG default parameters with \pthard $>3\gevc$.  The Monte-Carlo predictions 
are divided into two categories: charged particles that arise from the break-up of the beam and target (beam-beam remnants); 
and charged particles that arise from the outgoing jets plus initial and final-state radiation (hard scattering component).  
HERWIG $6.4$ has improved modeling of the parton-showers (modified leading-log), whereas ISAJET simply uses the leading-log 
approximation.  The modified leading-log takes into account the angle-ordering of the shower that is indicated by higher 
order corrections.  Clearly, the hard-scattering component of HERWIG does a much better job in describing the data than does the 
hard-scattering component of ISAJET.   However, it became clear that the beam-beam remnants of HERWIG were too soft to describe the CDF Run 1 
UE data.  To describe the data one needs to include MPI.

\begin{table}[htbp]
\caption{\footnotesize 
Parameters for several PYTHIA 6.2 tunes.  Tune A and Tune B are CDF Run 1 UE tunes.  Tune DW, D6, and DWT 
are CDF Run 2 tunes which fit the Run 2 UE data and fit the Run 1 $Z$-boson \pt\ distribution. 
Tune DW and Tune DWT are identical at $1.96\tev$ but have a different energy dependence.  Tune D6 is similar to 
Tune DW but uses CTEQ6L.
}
\label{table1}
\begin{tabular}{||c|c|c|c|c|c||} \hline \hline
 {\footnotesize\bf Parameter}  & {\bf Tune A}  & {\bf Tune B}  & {\bf Tune DW} & {\bf Tune D6} & {\bf Tune DWT}\\ \hline\hline
  {\footnotesize PDF} 	& CTEQ5L & CTEQ5L & CTEQ5L & CTEQ6L & CTEQ5L\\
  {\footnotesize MSTP(81)}	& $1$ & $1$ & $1$ & $1$ & $1$ \\
  {\footnotesize MSTP(82)}	& $4$ & $4$ & $4$ & $4$ & $4$ \\
  {\footnotesize PARP(82)}	& $2.0$ & $1.9$ & $1.9$ & $1.8$ & $1.9409$ \\
  {\footnotesize PARP(83)}	& $0.5$ & $0.5$ & $0.5$ & $0.5$ & $0.5$ \\
  {\footnotesize PARP(84)}	& $0.4$ & $0.4$ & $0.4$ & $0.4$ & $0.4$ \\
  {\footnotesize PARP(85)}	& $0.9$ & $0.9$ & $1.0$ & $1.0$ & $1.0$ \\
  {\footnotesize PARP(86)}	& $0.95$ & $0.95$ & $1.0$ & $1.0$ & $1.0$ \\
  {\footnotesize PARP(89)}	& $1800$ & $1800$ & $1800$ & $1800$ & $1960$\\
  {\footnotesize PARP(90)}	& $0.25$ & $0.25$ & $0.25$ & $0.25$ & $0.16$ \\
  {\footnotesize PARP(62)}	& $1.0$ & $1.0$ & $1.25$ & $1.25$ & $1.25$ \\
  {\footnotesize PARP(64)}	& $1.0$ & $1.0$ & $0.2$ & $0.2$ & $0.2$ \\
  {\footnotesize PARP(67)}	& $4.0$ & $1.0$ & $2.5$ & $2.5$ & $2.5$ \\
  {\footnotesize MSTP(91)}	& $1$ & $1$ & $1$ & $1$ & $1$ \\
  {\footnotesize PARP(91)}	& $1.0$ & $1.0$ & $2.1$ & $2.1$ & $2.1$ \\
  {\footnotesize PARP(93)}	& $5.0$ & $5.0$ & $15.0$ & $15.0$ & $15.0$ \\ \hline\hline
\end{tabular}
\end{table}

Figure~3 shows the first attempts to tune the parameters of PYTHIA $6.2$ to fit the CDF Run 1 UE data and Table 1 give the value 
of the parameters for the some of the CDF tunes. The perturbative $2$-to-$2$ parton-parton differential cross section diverges 
like  $1/\hat p_T^4$, where $\hat p_T$ is the transverse momentum of the outgoing parton in the parton-parton center-of-mass frame.  
PYTHIA regulates this cross section by including a smooth cut-off $p_{T0}$ as follows: 
$1/\hat p_T^4\rightarrow 1/(\hat p_T^2+p_{T0}^2)^2$.  This approaches the perturbative result 
for large scales and is finite as $\hat p_T\rightarrow0$.  The primary hard scattering processes and the MPI are regulated in the same way with the 
one parameter $p_{T0}=$ PARP($82$).  This parameter governs the amount of MPI in the event.  Smaller values of $p_{T0}$ results 
in more MPI due to a larger MPI cross-section.  PARP($67$) sets the high \pt\ scale for initial-state radiation in PYTHIA $6.2$.  
It determines the maximal parton virtuality allowed in time-like showers.  The larger the value of PARP($67$) the more initial-state 
radiation in the event.  Tune A has more initial-state radiation and less MPI and Tune B has less initial-state radiation and 
slightly more MPI.  However, as can be seen in Fig.~3 one cannot discriminate between them by simply looking at the activity 
in the \TR\ region.   

The CDF studies indicated that $p_{T0}$ is around $2\gevc$ at $1.96\tev$.  However, this cut-off is expected depend on the 
center-of-mass energy of the hadron-hadron collision, $E_{cm}$. PYTHIA parameterizes this energy dependence as follows:
$p_{T0}(E_{cm})=(E_{cm}/E_0)^\epsilon$, where $E_0=$ PARP($89$) is the reference energy and the parameter $\epsilon=$ PARP($90$) 
determines the energy dependence.  Fig.~4 shows how the PARP($90$) parameter for Tune A was determined. I determined the value 
of $\epsilon=$ 0.25 by comparing the UE activity in the CDF data at $1.8\tev$ and $630\gev$ \cite{cdf630}.  

Figure~5 shows that Tune A does not fit the CDF Run 1 Z-boson \pt\ distribution very well \cite{cdfzpt}.  PYTHIA Tune AW was adjusted to 
fit the Z-boson \pt\ distribution as well as the underlying event at the Tevatron. The UE activity of Tune A and Tune AW are 
nearly identical.  PYTHIA Tune DW is very similar to Tune AW except Tune DW has PARP($67$) $=2.5$, which is the preferred value 
determined by the D\O\ Collaboration in fitting their dijet $\Delta\phi$ distribution \cite{d0pub}.  

The MPI tune depends on the choice of parton distribution function (PDF).  One must choose a PDF and then tune to fit the UE.  
Tune A, B, AW, and DW use CTEQ5L.  Tune D6 is similar to tune DW except it uses CTEQ6L as the PDF.  Note that in changing 
from CTEQ5L to CTEQ6L, $p_{T0}=$ PARP($82$) decreased by a factor of $1.8/1.9\approx0.95$ in order to get the same UE activity.

PYTHIA Tune A, AW, DW, and D6 use $\epsilon=$ PARP($90$) $=0.25$, which is much different than the PYTHIA $6.2$ default value 
of $0.16$.  Tune DWT uses the default value of $0.16$.  Tune DW and Tune DWT are identical at $1.96\tev$, but Tune DW and DWT 
extrapolate to other energies differently. Tune DWT produces more activity in the UE at energies above the Tevatron than does 
Tune DW, but predicts less activity than Tune DW in the UE at energies below the Tevatron as shown in Fig.~6.  The data from 
the STAR Collaboration at $0.2\tev$ at RHIC data \cite{star} favor the energy dependence of Tune DW and rule out the energy dependence of Tune DWT. 

Fig.~7 shows the extrapolations of PYTHIA Tune A, Tune DW, Tune DWT, Tune S320, Tune P329, and pyATLAS to the LHC ($14\tev$).  
Tune pyATLAS is the original ATLAS tune that used the default parameter of PARP($90$) $=0.16$. Both Tune DWT and the old pyATLAS 
tune are ruled out by the RHIC UE data \cite{star}.  Tune S320 is the original Perugia0 tune from Peter Skands \cite{skands1} and Tune P329 is 
a ``professor" tune from Hendrik Hoeth.  In November of 2009 Tune DW, Tune S320, and Tune P329 seemed to be converging 
on the same predictions for the LHC.  I began to feel that we could make accurate LHC predictions with some confidence.  
Fig.~8 shows the charged particle density in the \TR\ region at 7 TeV as defined by PTmax, \CJone, and 
the muon-pair in Drell-Yan production as predicted from PYTHIA Tune DW.  The density of charged particles in the \TR\ region 
goes to zero as PTmax or \ptcj\ go to zero due to kinematics.  If PTmax is equal to zero then there are no charged particles 
anywhere in the $\eta$ region considered.  Similarly for \ptcj.  However, if the $PT$(muon-pair) goes to zero there is 
still the hard scale of the mass of the muon-pair and, hence, the charge particle density is not zero at $PT$(muon-pair) $=0$.

Figure~9 show the center-of-mass energy dependence of the charged particle density in the \TR\ region predicted by PYTHIA Tune DW.  
The height of the ``plateau" in the \TR\ region does not increase linearly with the center-of-mass energy.  For energies 
above the Tevatron it increases more like a straight line on a log plot (or a small power of $E_{cm}$).  The UE activity is 
predicted by PYTHIA Tune DW to increase by about a factor of two in going from $900\gev$ to $7\tev$ and then to 
increase by only about $20\%$ in going from $7\tev$ to $14\tev$.   

\begin{figure}[htbp]
\begin{center}
\includegraphics[scale=0.75]{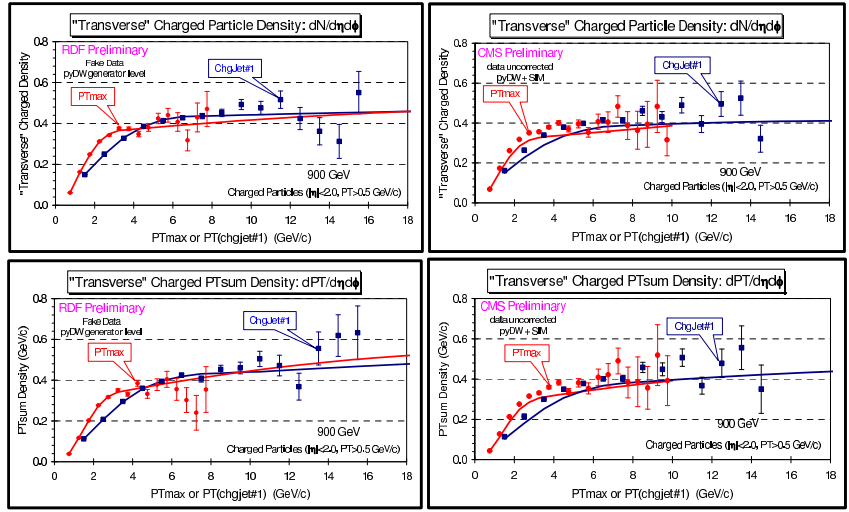}
\caption{\footnotesize
(\textit{left column}) Fake data at $900\gev$ on the transverse charged particle density (\textit{top left}) and the transverse 
charged PTsum density (\textit{bottom left}) as defined by the leading charged particle (PTmax) and the leading charged 
particle jet (\ptcj) for charged particles with \ptlcut and \etacut.  The fake data (from PYTHIA Tune DW) are generated 
at the particle level assuming $0.5$ M min-bias events at $900\gev$ ($361,595$ events in the plot). 
(\textit{right column}) Early CMS data \cite{cmsue1} at $900\gev$ on the transverse charged particle density 
(\textit{top right}) and the transverse charged PTsum density (\textit{bottom right}) as defined by the leading charged 
particle (PTmax) and the leading charged particle jet (\ptcj) for charged particles with \ptlcut and \etacut.  The data are 
uncorrected and compared with PYTHIA Tune DW after detector simulation ($216,215$ events in the plot).
}
\end{center}
\label{Zakopane_fig10}
\end{figure}

\begin{figure}[htbp]
\begin{center}
\includegraphics[scale=0.75]{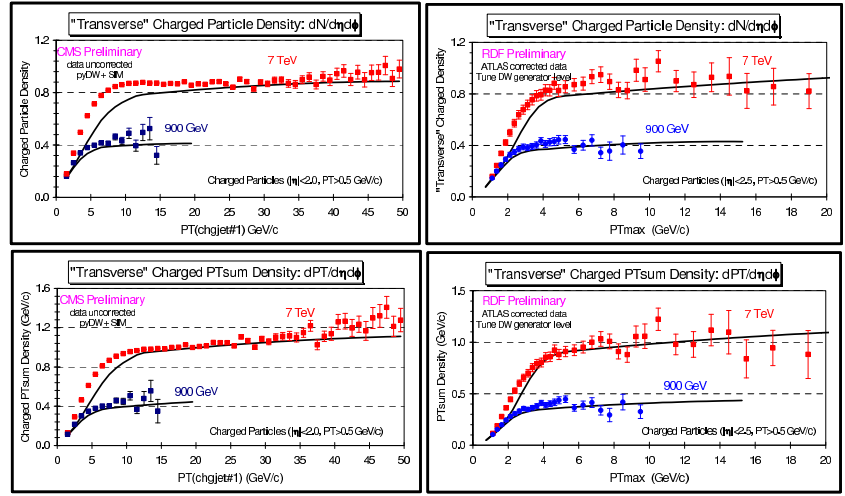}
\caption{\footnotesize
(\textit{left column}) Early CMS data at $900\gev$ and $7\tev$ \cite{cmsue2} on the transverse charged particle 
density (\textit{top left}) and the transverse charged PTsum density (\textit{bottom left}) as defined by the 
leading charged particle jet (\CJone) for charged particles with \ptlcut and \etacut.  The data are uncorrected 
and compared with PYTHIA Tune DW after detector simulation.  (\textit{right column}) Early ATLAS data \cite{atlas1} at 
$900\gev$ and $7\tev$ on the transverse charged particle density (\textit{top right}) and the 
transverse charged PTsum density (\textit{bottom right}) as defined by the leading charged particle (PTmax) for 
charged particles with \ptlcut and $|\eta|<2.5$.  The data are corrected and compared with PYTHIA Tune DW at 
the generator level.
}
\end{center}
\label{Zakopane_fig11}
\end{figure}

\begin{figure}[htbp]
\begin{center}
\includegraphics[scale=0.75]{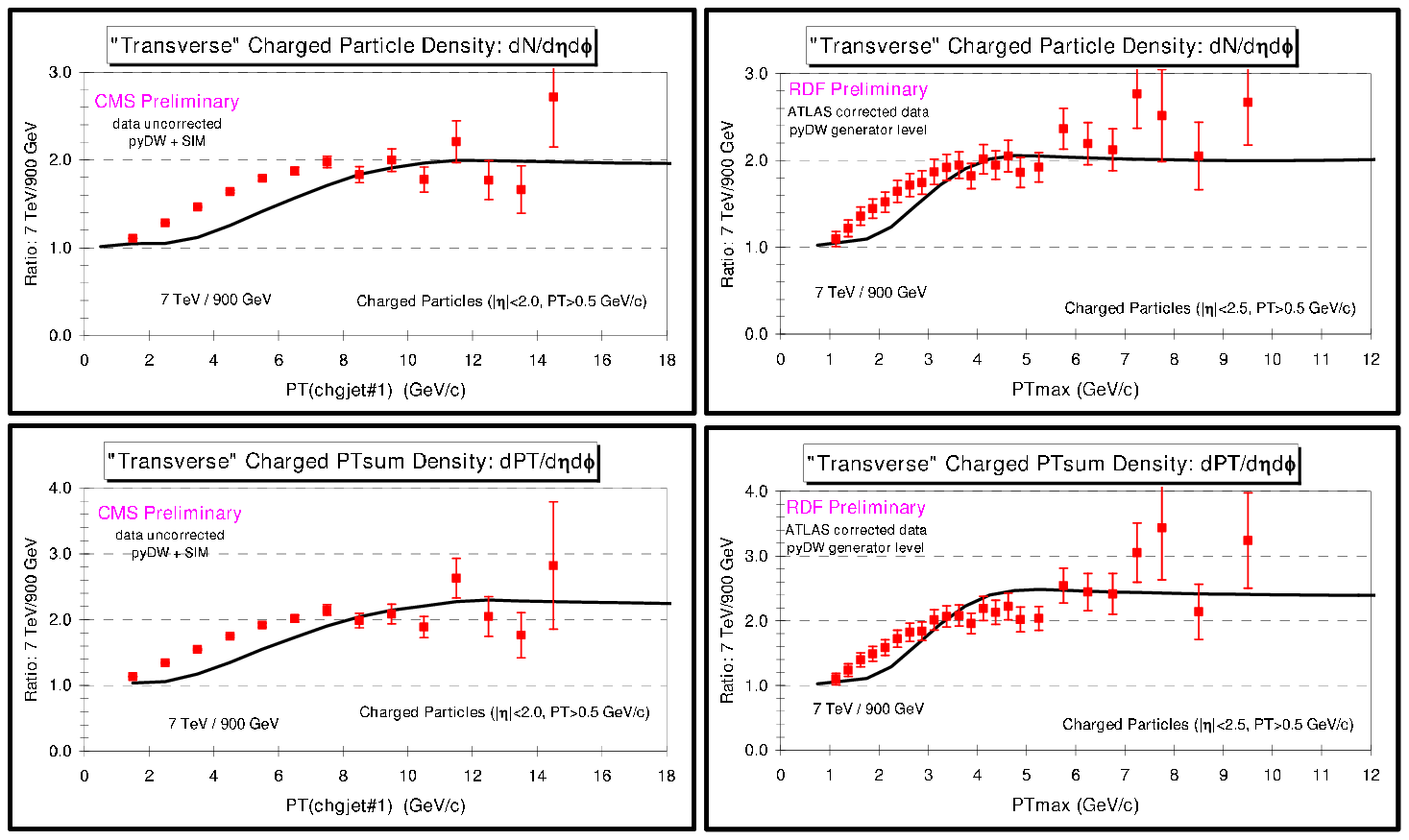}
\caption{\footnotesize
(\textit{left column}) Early CMS data on the ratio of $7\tev$ and $900\gev$ ($7\tev$ divided by $900\gev$ from Fig.~11) for 
the transverse charged particle density (\textit{top left}) and the transverse charged PTsum density (\textit{bottom left}) 
as defined by the leading charged particle jet (\CJone) for charged particles with \ptlcut and \etacut.  The data are uncorrected 
and compared with PYTHIA Tune DW after detector simulation.  (\textit{right column}) ATLAS data on the ratio of $7\tev$ and $900\gev$ 
from Fig. 11 for the transverse charged particle density (\textit{top right}) and the transverse charged PTsum density 
(\textit{bottom right}) as defined by the leading charged particle (PTmax) for charged particles with \ptlcut and $|\eta|<2.5$.  
The data are corrected and compared with PYTHIA Tune DW at the generator level.
}
\end{center}
\label{Zakopane_fig12}
\end{figure}

\begin{figure}[htbp]
\begin{center}
\includegraphics[scale=0.75]{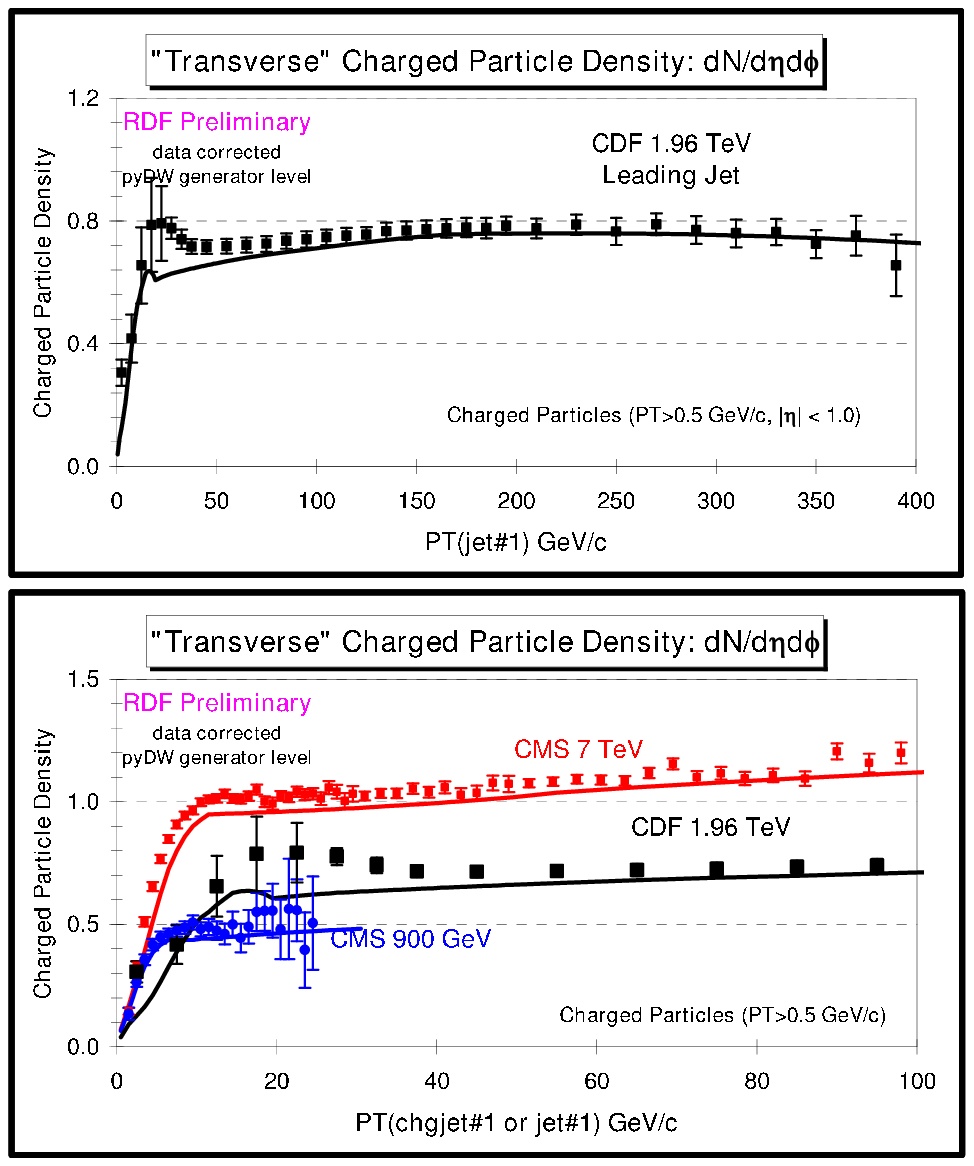}
\caption{\footnotesize
(\textit{top}) CDF data at $1.96\tev$ from Ref. \cite{cdfue2} on the charged particle density (\ptlcut, \etacdf) in the \TR\ region as 
defined by the leading calorimeter jet, \Jone, as a function of \ptone\ compared with PYTHIA Tune DW.  Also compares the CDF 
data at $1.96\tev$ with the recent CMS data \cite{cmsue3} at $900\gev$ and $7\tev$ (\textit{bottom}) on the \TR\ charged particle 
density (\ptlcut, \etacut) in the \TR\ region as defined by the leading charged particle jet, \CJone, as a function 
of \ptcj.  The data are corrected to the particle level and compared with PYTHIA Tune DW at the generator level.  
}
\end{center}
\label{Zakopane_fig13}
\end{figure}

\section{PYTHIA 6.2 Tune DW and the LHC UE Data}

The left column of Figure~10 shows two plots that I presented at the MB\&UE CMS Workshop at CERN on November 6, 2009 before we 
had LHC data.  The plots show generator level predictions of PYTHIA Tune DW at $900\gev$ for the transverse charged particle 
density and the transverse charged PTsum density as defined by the leading charged particle (PTmax) and the leading charged 
particle jet (\ptcj) for charged particles with \ptlcut\ and \etacut. The plots also show fake data at $900\gev$ generated 
from PYTHIA Tune DW assuming $500,000$ MB events ($361,595$ events in the plot).  The fake data agree perfectly with Tune DW 
since it was generated from Tune DW!  This is what I expected the data to look like if CMS received $500,000$ MB triggers at $900\gev$.
The right column of Figure~10 shows the data CMS collected at the LHC during the commissioning period of December 2009 \cite{cmsue1}.  
The data are uncorrected and compared with PYTHIA Tune DW after detector simulation ($216,215$ events in the plot).   
CMS did not quite get $500,000$ MB triggers, but we got enough to get a first look at the underlying event activity at $900\gev$.  
PYTHIA Tune DW did a fairly good job in describing the features of this data, but it does not fit the data perfectly.  
It does not fit the real data quite as well as it fits the fake data!  However, we saw roughly what we expected to see.

Figure~11 shows early CMS \cite{cmsue2} and ATLAS \cite{atlas1} data at $900\gev$ and $7\tev$ on the transverse charged particle density 
and the transverse charged PTsum density compared with the predictions of PYTHIA Tune DW.  Here CMS useed the leading charged 
particle jet (\CJone) to define the transverse region and ATLAS used the leading charged particle, PTmax.  The ATLAS data 
are corrected to the particle level and compared with Tune DW at the generator level.  The CMS data are uncorrected and 
compared with Tune DW after detector simulation.  Tune DW predicted about the right amount of activity in the 
``plateau", but does not fit the low \pt\ rise very well.  Figure~12 shows early CMS and ATLAS data on the ratio between $7\tev$ and 
$900\gev$ ($7\tev$ divided by $900\gev$) for the transverse charged particle density and the transverse charged PTsum density 
compared with PYTHIA Tune DW.   Tune DW predicted that the transverse charged particle density would increase by about a 
factor of two in going from $900\gev$ to $7\tev$ and that the transverse PTsum density would have a slightly larger increase.  
Both these predictions are seen in the data, although Tune DW does not fit very well the energy dependence of the low \pt\ 
approach to the ``plateau".

Figure~13 shows the CDF data at $1.96\tev$ \cite{cdfue2} on the charged particle density in the \TR\ region as defined by the leading 
calorimeter jet, \Jone, as a function of \ptone\ together with the recent CMS data \cite{cmsue3} at $900\gev$ and $7\tev$ on the charged particle 
density in the \TR\ region as defined by the leading charged particle jet, \CJone, as a function of \ptcj\ compared with 
PYTHIA Tune DW.  I would say that the agreement at all three energies is fairly good.  Tune DW, however, is not a perfect 
fit to the LHC UE data.  It does not fit the Tevatron data perfectly either!  We expect a lot from the QCD Monte-Carlo 
models.  We want them to fit perfectly which is, of course, not always possible.  

\begin{figure}[htbp]
\begin{center}
\includegraphics[scale=0.75]{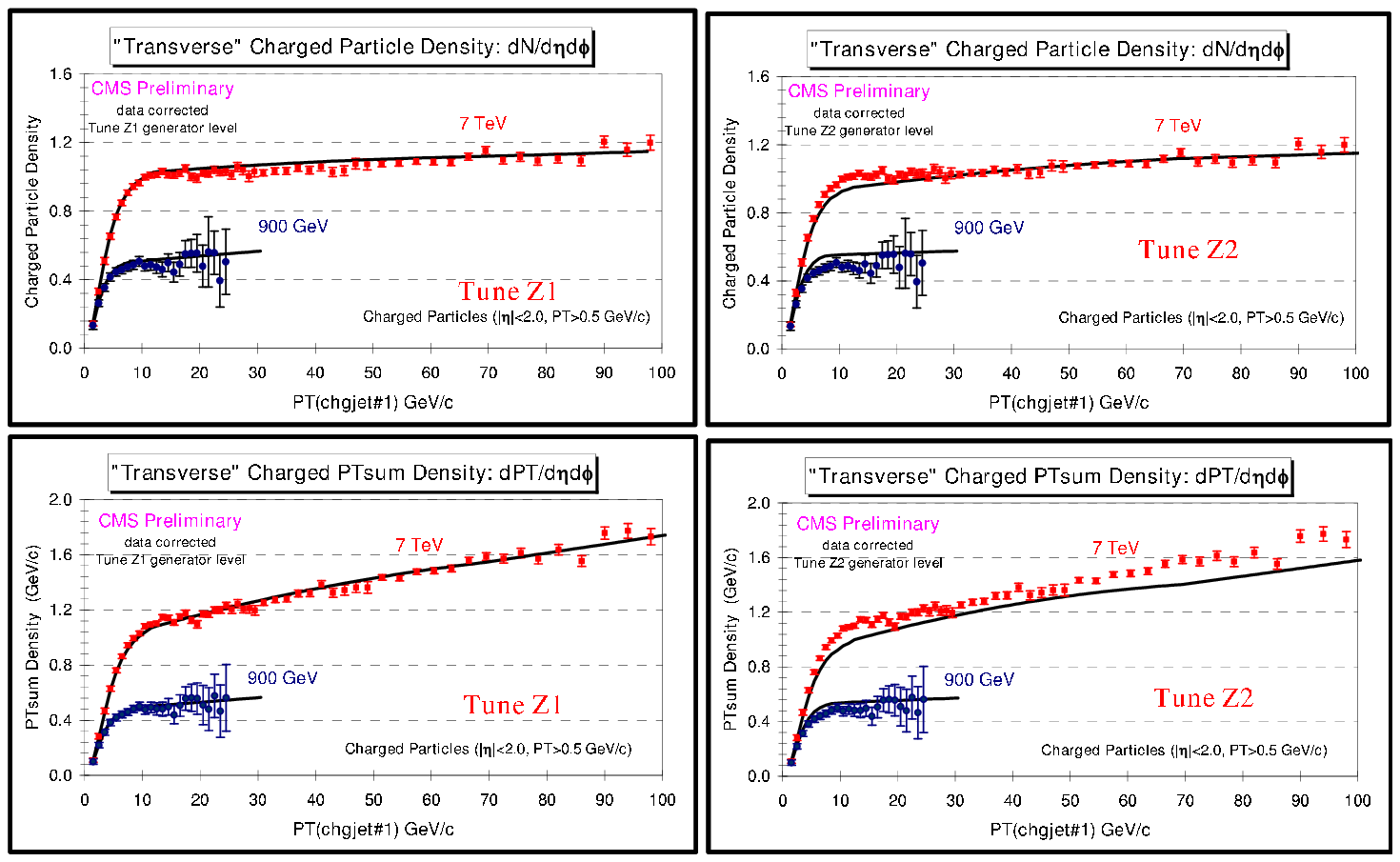}
\caption{\footnotesize
(\textit{left column}) CMS data at $900\gev$ and $7\tev$ \cite{cmsue3} on the transverse charged particle density (\textit{top left}) 
and the transverse charged PTsum density (\textit{bottom left}) as defined by the leading charged particle jet (\CJone) for 
charged particles with \ptlcut and \etacut.  The data are corrected to the particle level and compared and with 
PYTHIA Tune Z1 at the generator level.  (\textit{right column}) CMS preliminary data at $900\gev$ and $7\tev$ on the transverse 
charged particle density (\textit{top left}) and the transverse charged PTsum density (\textit{bottom left}) as defined by the 
leading charged particle jet (\CJone) for charged particles with \ptlcut and \etacut.  The data are corrected to the particle 
level and compared and with PYTHIA Tune Z2 at the generator level.
}
\end{center}
\label{Zakopane_fig14}
\end{figure}

\begin{figure}[htbp]
\begin{center}
\includegraphics[scale=0.75]{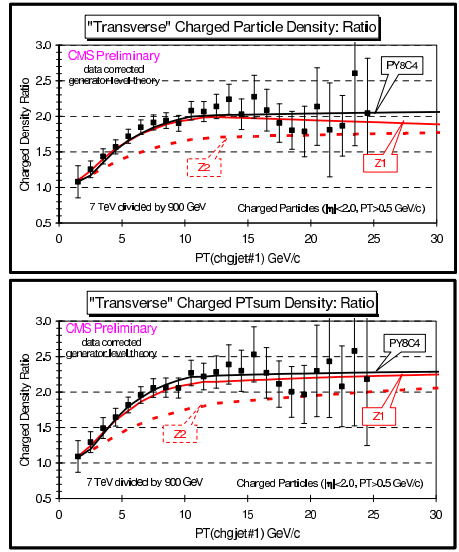}
\caption{\footnotesize
CMS data on the ratio of $7\tev$ and $900\gev$ ($7\tev$ divided by $900\gev$ from Fig. 14) for the transverse charged particle 
density (\textit{top}) and the transverse charged PTsum density (\textit{bottom}) as defined by the leading charged particle 
jet (\CJone) for charged particles with \ptlcut and \etacut.  The data are corrected to the particle level and compared and 
with PYTHIA Tune Z1, Tune Z2, and PYTHIA 8 Tune 4C \cite{corke} at the generator level.
}
\end{center}
\label{Zakopane_fig15}
\end{figure}

\begin{table}[htbp]
\caption{\footnotesize 
PYTHIA 6.4 parameters for the ATLAS Tune AMBT1 and the CMS UE Tune Z1 and Z2.  
Parameters not shown are set to their defuult value.
}
\label{table2}
\begin{tabular}{||l|c|c|c||} \hline \hline
 {\bf Parameter}  & {\bf Tune Z1}  & {\bf Tune Z2}  & {\bf AMBT1} \\ \hline\hline
 {\footnotesize PDF - Parton Distribution Function} 	& CTEQ5L & CTEQ6L & LO* \\
 {\footnotesize PARP(82) - MPI Cut-off}			& $1.932$ & $1.832$ & $2.292$ \\
 {\footnotesize PARP(89) - Reference energy, E$_0$}	& $1800$ & $1800$ & $1800$ \\
 {\footnotesize PARP(90) - MPI Energy Extrapolation}	& $0.275$ & $0.275$ & $0.25$ \\
 {\footnotesize PARP(77) - CR Suppression}			& $1.016$ & $1.016$ & $1.016$ \\
 {\footnotesize PARP(78) - CR Strength}			& $0.538$ & $0.538$ & $0.538$  \\
 {\footnotesize PARP(80) - Probability colored parton from BBR}	& $0.1$ & $0.1$ & $0.1$ \\
 {\footnotesize PARP(83) - Matter fraction in core}	& $0.356$ & $0.356$ & $0.356$ \\
 {\footnotesize PARP(84) - Core of matter overlap}		& $0.651$ & $0.651$ & $0.651$ \\
 {\footnotesize PARP(62) - ISR Cut-off}			& $1.025$ & $1.025$ & $1.025$ \\
 {\footnotesize PARP(93) - primordial kT-max} 		& $10.0$ & $10.0$ & $10.0$ \\ 
 {\footnotesize MSTP(81) - MPI, ISR, FSR, BBR model}	& $21$ & $21$ & $21$ \\
 {\footnotesize MSTP(82) - Double gaussion matter distribution}	& $4$ & $4$ & $4$ \\
 {\footnotesize MSTP(91) - Gaussian primordial kT}		& $1$ & $1$ & $1$ \\
 {\footnotesize MSTP(95) - strategy for color reconnection}	& $6$ & $6$ & $6$ \\ \hline\hline
\end{tabular}
\end{table}

\begin{figure}[htbp]
\begin{center}
\includegraphics[scale=0.75]{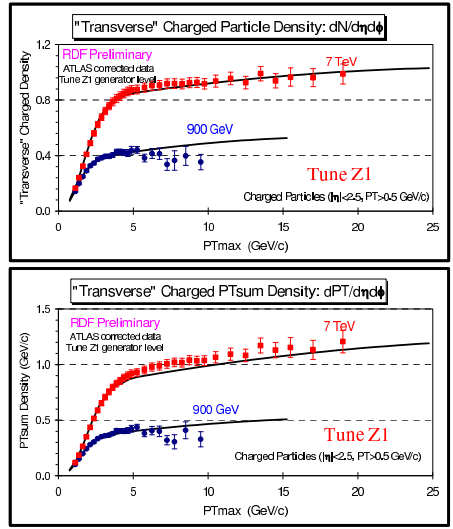}
\caption{\footnotesize
ATLAS data at $900\gev$ and $7\tev$ \cite{atlas2} on the transverse charged particle density (\textit{top}) and the transverse charged 
PTsum density (\textit{bottom}) as defined by the leading charged particle, PTmax, as a function of PTmax for charged particles 
with \ptlcut and $|\eta|<2.5$.  The data are corrected to the particle level and compared with PYTHIA Tune Z1 at the generator level.
}
\end{center}
\label{Zakopane_fig16}
\end{figure}

\begin{figure}[htbp]
\begin{center}
\includegraphics[scale=0.75]{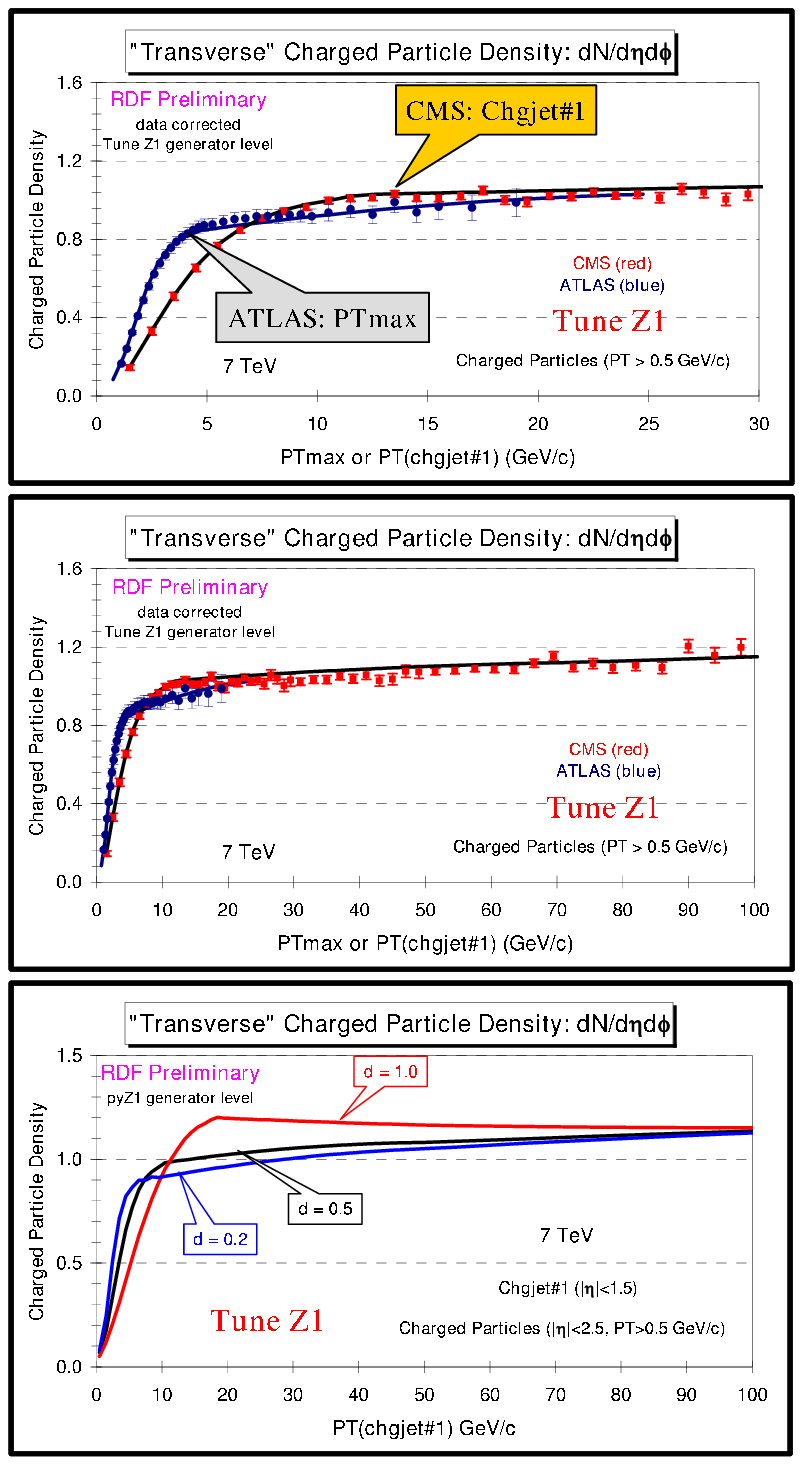}
\caption{\footnotesize
CMS data from Fig.~14 on the charged particle density in the \TR\ region as defined by the leading charged particle jet, \CJone, versus 
\ptcj\ up to $30\gevc$ (\textit{top}) and up to $100\gevc$ (\textit{middle}) compared with the ATLAS data from Fig.~16 on the charged 
particle density in the \TR\ region as defined by the leading charged particle, PTmax, versus PTmax. The data are corrected to the 
particle level and compared with PYTHIA Tune Z1 at the generator level. (\textit{bottom}) Dependence of the transverse charged 
particle density on the charged particle jet radius as predicted by PYTHIA Tune Z1.  Charged particle jets are constructed 
using the Anti-KT algorithm with $d=0.2$, $0.5$, and $1.0$.  The charged particles have \ptlcut and $|\eta|<2.5$ and 
the leading charged particle jet is restricted to be in the region $|\eta({\rm chgjet}\#1)|<1.5$.
}
\end{center}
\label{Zakopane_fig17}
\end{figure}

\begin{figure}[htbp]
\begin{center}
\includegraphics[scale=0.75]{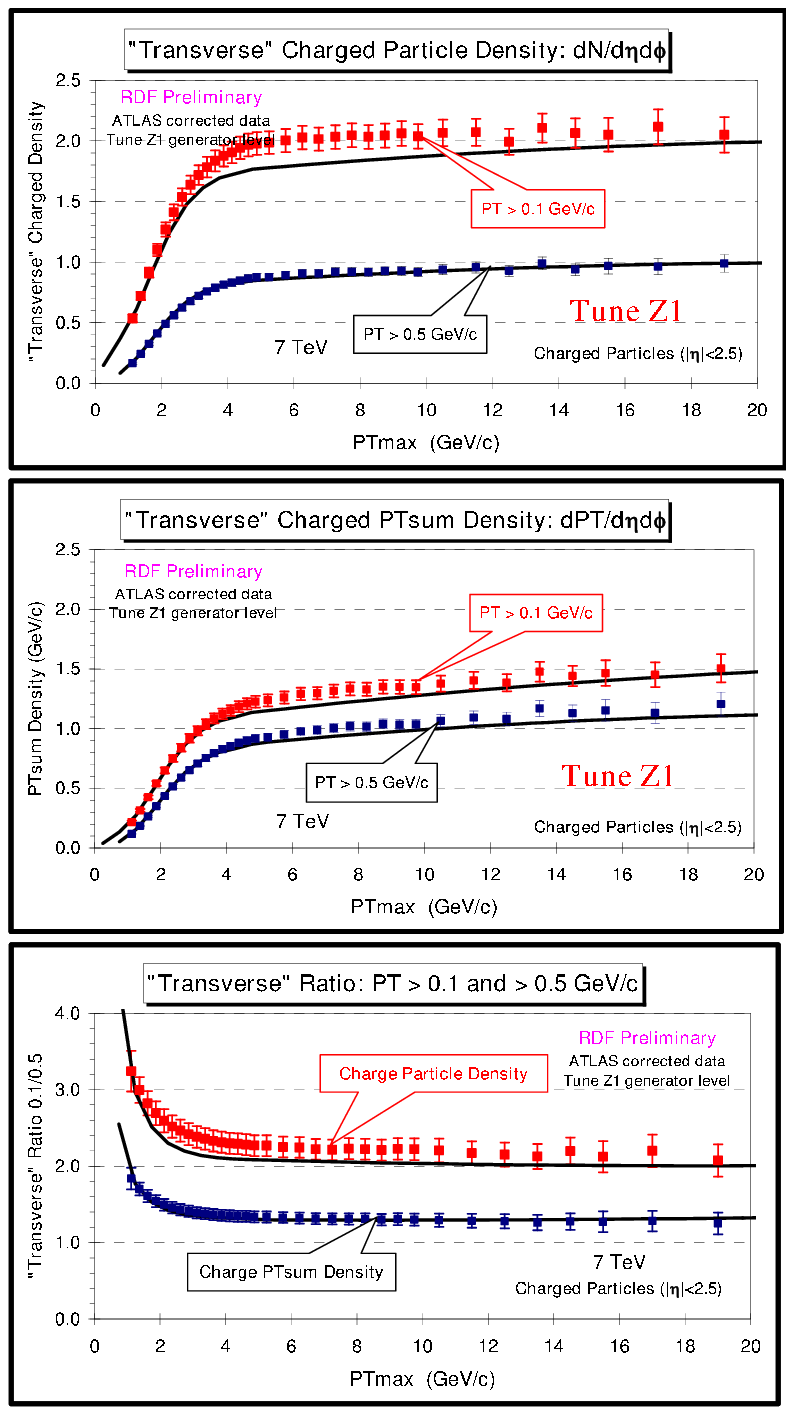}
\caption{\footnotesize
ATLAS data at $7\tev$ \cite{atlas2} on the transverse charged particle density (\textit{top}) and the transverse charged PTsum 
density (\textit{middle}) as defined by the leading charged particle, PTmax, as a function of PTmax for charged particles 
with \ptlcut and $|\eta|<2.5$ and for charged particles with $p_T> 0.1\gevc$ and $|\eta|<2.5$.  The data are corrected to 
the particle level and compared with PYTHIA Tune Z1 at the generator level.  (\textit{bottom}) Ratio of the ATLAS data 
with $p_T> 0.1\gevc$ and \ptlcut\ ($100\mevc$ cut divided by $500\mevc$ cut) for the transverse charged particle density 
and the charged PTsum density as defined by the leading charged particle, PTmax, as a function of PTmax compared with PYTHIA Tune Z1.
}
\end{center}
\label{Zakopane_fig18}
\end{figure}

\begin{figure}[htbp]
\begin{center}
\includegraphics[scale=0.75]{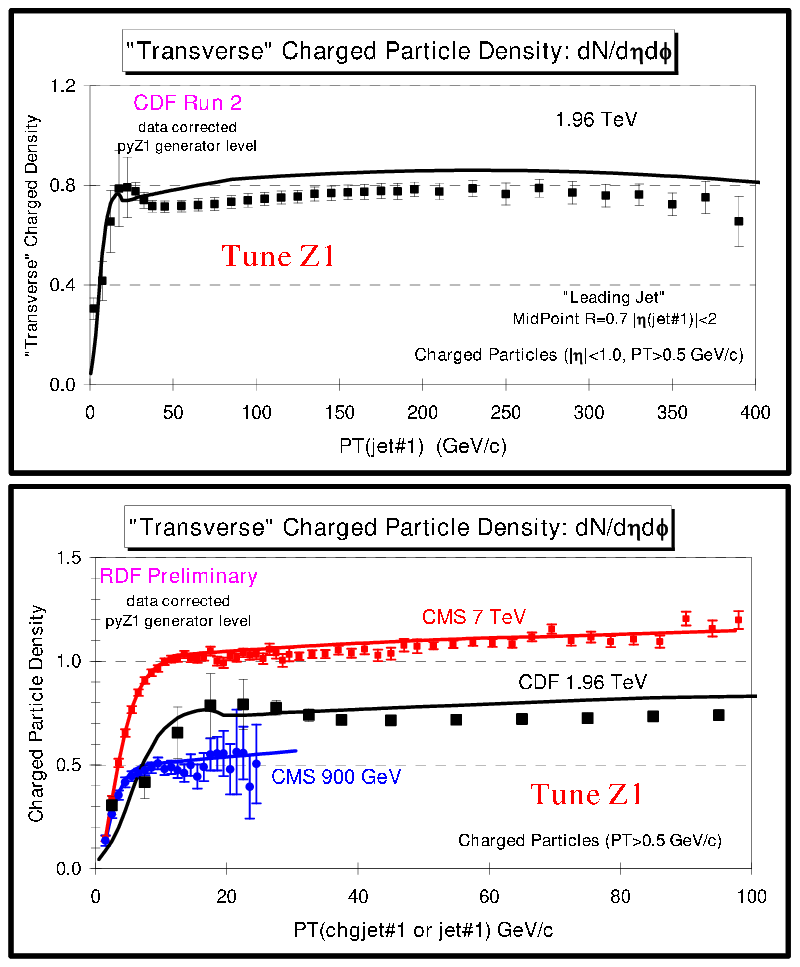}
\caption{\footnotesize
(\textit{top}) CDF data at $1.96\tev$ from Ref. \cite{cdfue2} on the charged particle density (\ptlcut, \etacdf) in the \TR\ region as 
defined by the leading calorimeter jet, \Jone, as a function of \ptone\ compared with PYTHIA Tune Z1.  Also compares the CDF 
data at $1.96\tev$ with the CMS data \cite{cmsue3} at $900\gev$ and $7\tev$ (\textit{bottom}) on the \TR\ charged particle 
density (\ptlcut, \etacut) in the \TR\ region as defined by the leading charged particle jet, \CJone, as a function 
of \ptcj.  The data are corrected to the particle level and compared with PYTHIA Tune Z1 at the generator level.  
}
\end{center}
\label{Zakopane_fig19}
\end{figure}

\begin{figure}[htbp]
\begin{center}
\includegraphics[scale=0.75]{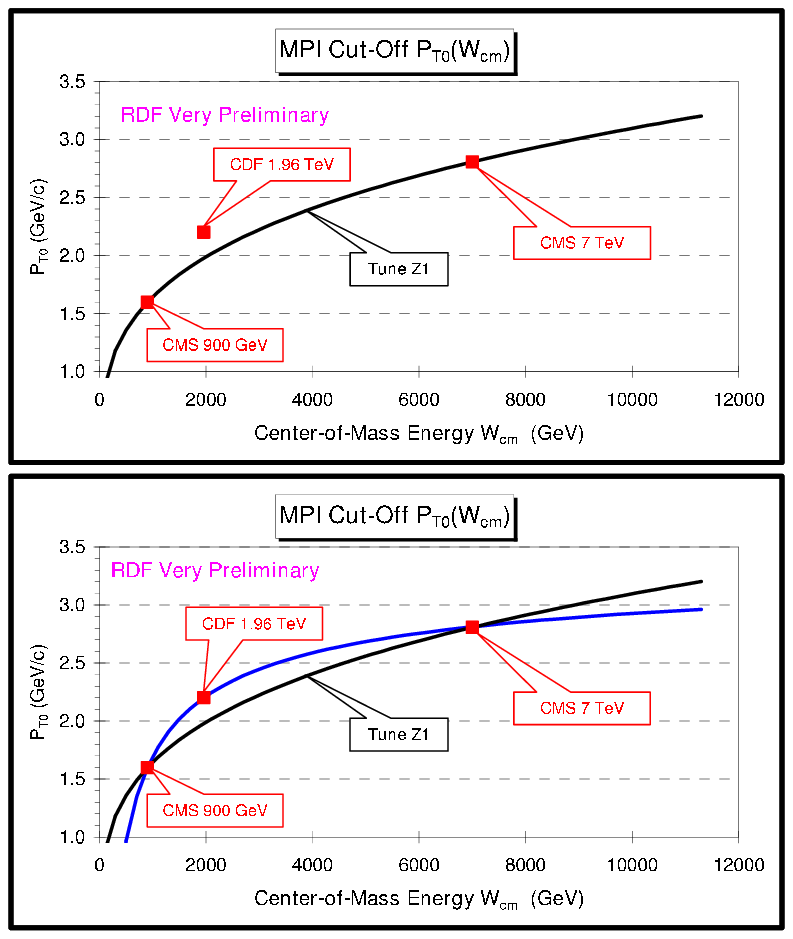}
\caption{\footnotesize
(\textit{top}) Shows the PYTHIA Tune Z1 $2$-to-$2$ hard scattering cut-off, $p_{T0}$, at $900\gev$ and $7\tev$ as determined 
by fitting the CMS UE data in Fig.~14 together with PYTHIA's functional form of 
$p_{T0}(W_{cm})=p_{T0}(W_{cm}/W_0)^\epsilon$ with $p_{T0}=$ PARP($82$) $= 1.932\gevc$, $\epsilon=$ PARP($90$) $=0.275$ 
and $W_0=1.8\tev$.  Also, shows the value of $p_{T0}$ at $1.96\tev$ that would fit better the CDF data in Fig.~19.  The PYTHIA 
functional form predicts a smaller cut-off at $1.96\tev$ resulting in a more active UE than observed in the CDF leading jet data 
in Fig.~19. (\textit{bottom})  Same as the top plot, but with an arbitrary functional form that extrapolates through the 
three energy points. 
}
\end{center}
\label{Zakopane_fig20}
\end{figure}

\section{PYTHIA 6.4 CMS UE Tune Z1 and Tune Z2}

Tune DW is a PYTHIA $6.2$ tune (Q$^2$-ordered parton showers, old MPI model) designed by me to fit the CDF underlying event 
data at $1.96\tev$.  Now that we have LHC data at $900\gev$ and $7\tev$ both ATLAS and CMS have LHC tunes.  The ATLAS 
Tune AMBT1 \cite{ambt1} is a PYTHIA $6.4$ tune ($p_T$-ordered parton showers, new MPI model) designed to fit the ATLAS LHC MB data 
for Nchg $\ge 6$ and \ptlcut\ (\ie ``diffraction suppressed MB").  They also included their underlying event data for 
PTmax $>5\gevc$, but the errors on the data are large in this region and hence their UE data did not have much influence 
on the resulting tune.  The ATLAS AMBT1 tune does significantly better at fitting the LHC ``diffraction suppressed MB" data, 
but does not do so well at fitting the LHC underlying event data.  I started with the ATLAS Tune AMBT1 and varied a few of 
the parameter to improve the fit to the CMS underlying event data at $900\gev$ and $7\tev$.  The parameters of the ATLAS 
Tune AMBT1 and the CMS UE Tune Z1 are shown in Table~2. 

Tune Z1 is a PYTHIA $6.4$ tune that uses the CTEQ5L PDF.  CDF wanted also a PYTHIA $6.4$ tune that uses the CTEQ6L PDF.  I know 
from my experience with Tune DW and Tune D6 (Table 1) that in going from CTEQ5L to CTEQ6L that I would have to decrease the 
value of $p_{T0}=$ PARP($82$), so I decreased it by a factor of $0.95$ (which is precisely the Tune D6 to Tune DW ratio) and produced 
Tune Z2.  The parameters of Tune Z2 are also shown in Table 2.  In my haste, I set  $\epsilon=$ PARP($90$) $=0.275$ for Tune Z2 
which is the same value that I deermined for Tune Z1.

Figure~14 shows the more recent CMS data at $900\gev$ and $7\tev$ \cite{cmsue3} on the transverse charged particle density and the transverse charged PTsum 
density as defined by the leading charged particle jet, \CJone. This data are corrected to the particle level and compared and 
with PYTHIA Tune Z1 and Tune Z2.  Tune Z1 does a much better job in describing the low \pt\ rise to the plateau than Tune DW. 
Tune Z2 does not describe the CMS UE data quite as well as Tune Z1.

Figure~15 shows CMS data on the ratio of $7\tev$ and $900\gev$ ($7\tev$ divided by $900\gev$) for the transverse charged particle 
density and the transverse charged PTsum density compared with PYTHIA Tune Z1, Tune Z2, and the PYTHIA 8 Tune 4C from 
Corke and Sj\"ostrand \cite{corke}.  Tune Z1 and Tune Z2 have the same value of PARP($90$) $=0.275$, however, Tune Z1 fits the energy dependence 
quite nicely while Tune Z2 does not.   In constructing Tune Z2, I forgot that the PDF also affects the energy dependence.  When I 
changed from CTEQ5L (Tune Z1) to CTEQ6L (Tune Z2) I should have also changed $\epsilon=$ PARP($90$) as well as PARP($82$).  The PYTHIA 8 
Tune 4C uses CTEQ6L but has $\epsilon=0.19$ and fits the energy dependence very nicely.  However, Tune 4C does not fit the LHC UE 
data at $900\gev$ and $7\tev$ as well as Tune Z1 does.

Figure~16 shows the latest ATLAS data \cite{atlas2} at $900\gev$ and $7\tev$ on the transverse charged particle density and the transverse charged 
PTsum density as defined by the leading charged particle, PTmax, for charged particles with \ptlcut\ and $|\eta|< 2.5$. The 
data are corrected to the particle level and compared with PYTHIA Tune Z1.  Tune Z1 describes very well both the CMS and ATLAS UE 
data. Fig.~17 compares the CMS data using \CJone\ with the ATLAS data which uses PTmax approach.  Tune Z1 describes the differences 
between the CMS \CJone\ and the ATLAS PTmax approach very well.  It is interesting that the activity in the ``plateau" of the \TR\ region 
is larger for the \CJone\ approach than it is for the PTmax analysis. Could it be that when one requires a charged particle 
jet with a certain value of \ptcj\ that you bias the UE to be more active, because a more active UE can contribute some \pt\ to the 
leading charged particle jet?  In an attempt to understand this, in Fig.~17 I looked at the dependence of the transverse charged 
particle density on the charged particle jet size (\ie radius) as predicted by PYTHIA Tune Z1.  I constructed charged particle 
jets using the Anti-KT algorithm with $d=0.2$, $0.5$, and $1.0$.  The charged particles have \ptlcut\ 
and $|\eta|< 2.5$ and the leading charged particle jet is restricted to be in the 
region $|\eta({\rm chgjet}\#1)|<1.5$.  For very narrow jets the UE ``plateau" is nearly the same as 
in the PTmax approach.  As the jets become larger in radius the UE ``plateau" becomes more active!   The object that is 
being used to define the \TR\ region can bias the UE to be more active.  Amazing! 

Figure~18 shows the recent ATLAS data at $7\tev$ \cite{atlas2} on the transverse charged particle density and the transverse charged PTsum density 
as defined by the leading charged particle, PTmax, for charged particles with $p_T>0.1\gevc$ and $|\eta|< 2.5$ compared with 
PYTHIA Tune Z1.  Fig.~18 also shows the ratio of the ATLAS data with $p_T>0.1\gevc$ and $p_T>0.5\gevc$ compared with 
PYTHIA Tune Z1.  All of the CDF UE measurements involved charged particles with \ptlcut.  This is the first look at the UE in 
the region below $500\mevc$ and there are a lot of soft particles!  The transverse charged particle density increases by 
about a factor of $2$ in going from $p_T>0.5\gevc$ to $p_T>0.1\gevc$.  Tune Z1 describes this increase better than 
Tune DW, however, Tune Z1 still does not have quite enough soft particles.   

Figure~19 shows the CDF data at $1.96\tev$ on the charged particle density in the \TR\ region as defined by the leading calorimeter 
jet, \Jone, as a function of \ptone\ together with the CMS data at $900\gev$ and $7\tev$ on the charged particle density in 
the "transverse" region as defined by the leading charged particle jet, \CJone, as a function of \ptcj\ compared with PYTHIA 
Tune Z1.  Fig.~19 shows the CMS PYTHIA $6.4$ Tune Z1 and Fig. 13 shows the CDF PYTHIA $6.2$ Tune DW.  Neither of the tunes 
describe perfectly all three energies.  Tune Z1 is in very good agreement with the UE data at $900\gev$ and $7\tev$ but is a 
little high at $1.96\tev$.  One can see this in Fig.~20 which shows the PYTHIA Tune Z1 $2$-to-$2$ hard scattering 
cut-off, $p_{T0}$, at $900\gev$ and $7\tev$ as determined by fitting the CMS UE data in Fig.~14 together with functional form of PYTHIA, 
$p_{T0}(W_{cm})=(W_{cm}/W_0)^\epsilon$, with $P_{T0}=$ PARP($82$) $=1.932\gev$, $\epsilon=$ PARP($90$) $=0.275$, 
and $W_0=1.8\tev$.  Here $W_{cm}=E_{cm}$ is the hadron-hadron center-of-mass energy.  Fig.~20 also shows the value of $p_{T0}$ at $1.96\tev$ that would fit better the CDF data in Fig.~19.  
The PYTHIA functional form predicts a smaller cut-off at $1.96\tev$ resulting in a more active UE than observed in the 
CDF ``leading jet" data in Fig.~19 \cite{skands2}.   I believe it is premature to consider other functional forms for $P_{T0}(W_{cm})$.  
I believe that we will find a PYTHIA tune that simultaneously describes $900\gev$, $1.96\tev$, and $7\tev$.  Remember the 
energy dependence of the UE depends not only on $\epsilon=$ PARP($90$), but also on the choice of PDF!

\begin{figure}[htbp]
\begin{center}
\includegraphics[scale=0.75]{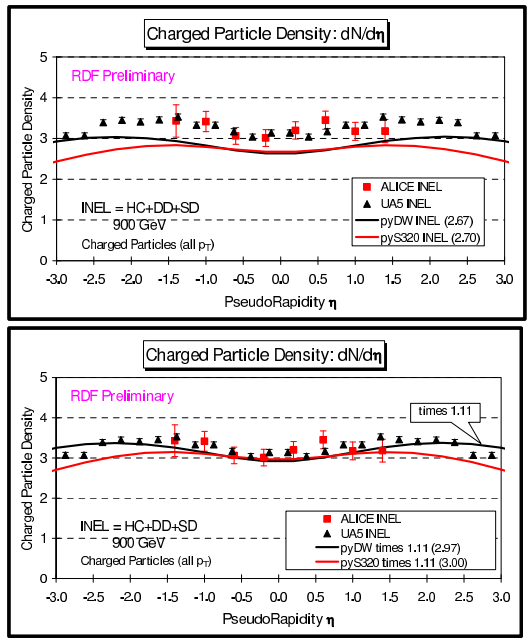}
\caption{\footnotesize
(\textit{top}) The inelastic (INEL) data from ALICE and UA5 at $900\gev$ \cite{alice1} on the charged particle density, $dN/d\eta$ (all \pt) 
compared with PYTHIA Tune DW and Tune S320. (\textit{bottom}) Same as the top plot except the Monte-Carlo model predictions have been 
multiplied by a factor of $1.11$.
}
\end{center}
\label{Zakopane_fig21}
\end{figure}

\begin{figure}[htbp]
\begin{center}
\includegraphics[scale=0.75]{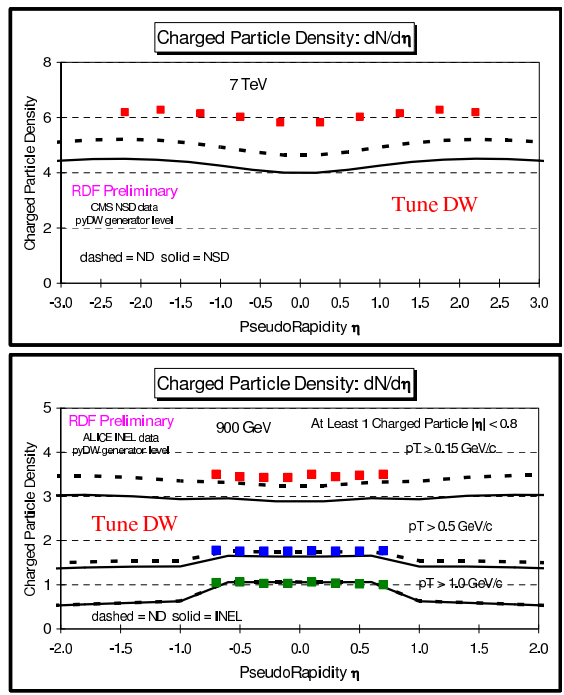}
\caption{\footnotesize
(\textit{top}) The non-single diffraction (NSD) data from CMS at $7\tev$ \cite{cmsmb1} on the charged particle density, $dN/d\eta$ (all \pt) 
compared with PYTHIA Tune DW.  The solid curve is NSD and the dashed curve is inelastic non-diffraction (ND) component. 
(\textit{bottom}) The inelastic (INEL) data from ALICE \cite{alice2} at $900\gev$ on the charged particle density, $dN/d\eta$, 
with $p_T>$ PTcut and at lease one charged particle with $p_T>$ PTcut and $|\eta|<0.8$ for PTcut $= 0.15\gevc$, $0.5\gevc$, and $1.0\gevc$ 
compared with PYTHIA Tune DW. The solid curve is the INEL and the dashed curve is inelastic non-diffraction (ND) component.
}
\end{center}
\label{Zakopane_fig22}
\end{figure}

\begin{figure}[htbp]
\begin{center}
\includegraphics[scale=0.6]{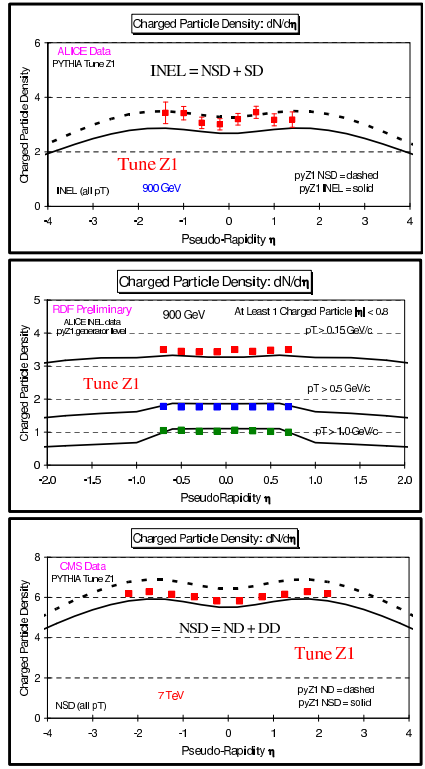}
\caption{\footnotesize
(\textit{top}) The inelastic (INEL) data from ALICE \cite{alice1} at $900\gev$ \cite{alice1} on the charged particle density, $dN/d\eta$ (all \pt) compared 
with PYTHIA Tune Z1. The solid curve is the INEL and the dashed curve is non-single diffraction (NSD) component. 
(\textit{middle}) The inelastic (INEL) data from ALICE \cite{alice2} at $900\gev$ on the charged particle density, $dN/d\eta$, 
with $p_T>$ PTcut and at lease one charged particle with $p_T>$ PTcut and $|\eta|<0.8$ for PTcut $= 0.15\gevc$, $0.5\gevc$, 
and $1.0\gevc$ compared with PYTHIA Tune Z1. (\textit{bottom}) The non-single diffraction (NSD) data from CMS \cite{cmsmb1} at $7\tev$ on 
the charged particle density, $dN/d\eta$ (all \pt) compared with PYTHIA Tune Z1.  The solid curve is NSD and the dashed 
curve is inelastic non-diffraction (ND) component.
}
\end{center}
\label{Zakopane_fig23}
\end{figure}

\begin{figure}[htbp]
\begin{center}
\includegraphics[scale=0.8]{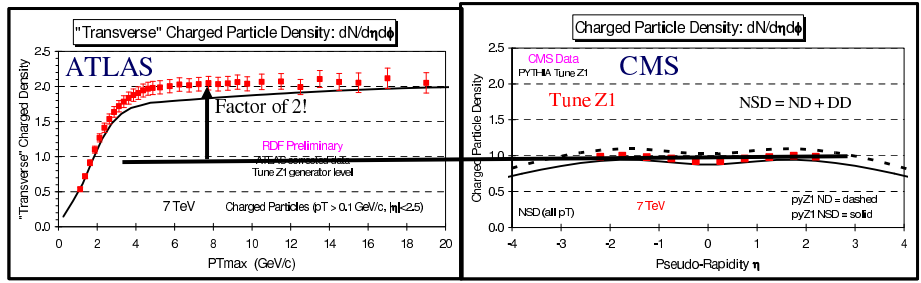}
\caption{\footnotesize
(\textit{right}) The non-single diffraction (NSD) data from CMS $7\tev$ \cite{cmsmb1} on the charged particle density, $dN/d\eta d\phi$ (all \pt) 
compared with PYTHIA Tune Z1.  The data and theory on $dN/d\eta$ in Fig.~23 has been divided by $2\pi$ to construct the number 
of particles per unit \etaphi.  (\textit{left}) ATLAS data from Fig.~18 at $7\tev$ on the charged particle density in 
the \TR\ region as defined by the leading charged particle, PTmax, as a function of PTmax for charged particles 
with $p_T > 0.1\gevc$ and $|\eta| < 2.5$ compared with PYTHIA Tune Z1.  The activity in the UE of a hard scattering 
process (\textit{left}) is a factor of two greater than in an average MB collision (\textit{right}).
}
\end{center}
\label{Zakopane_fig24}
\end{figure}

\begin{figure}[htbp]
\begin{center}
\includegraphics[scale=0.75]{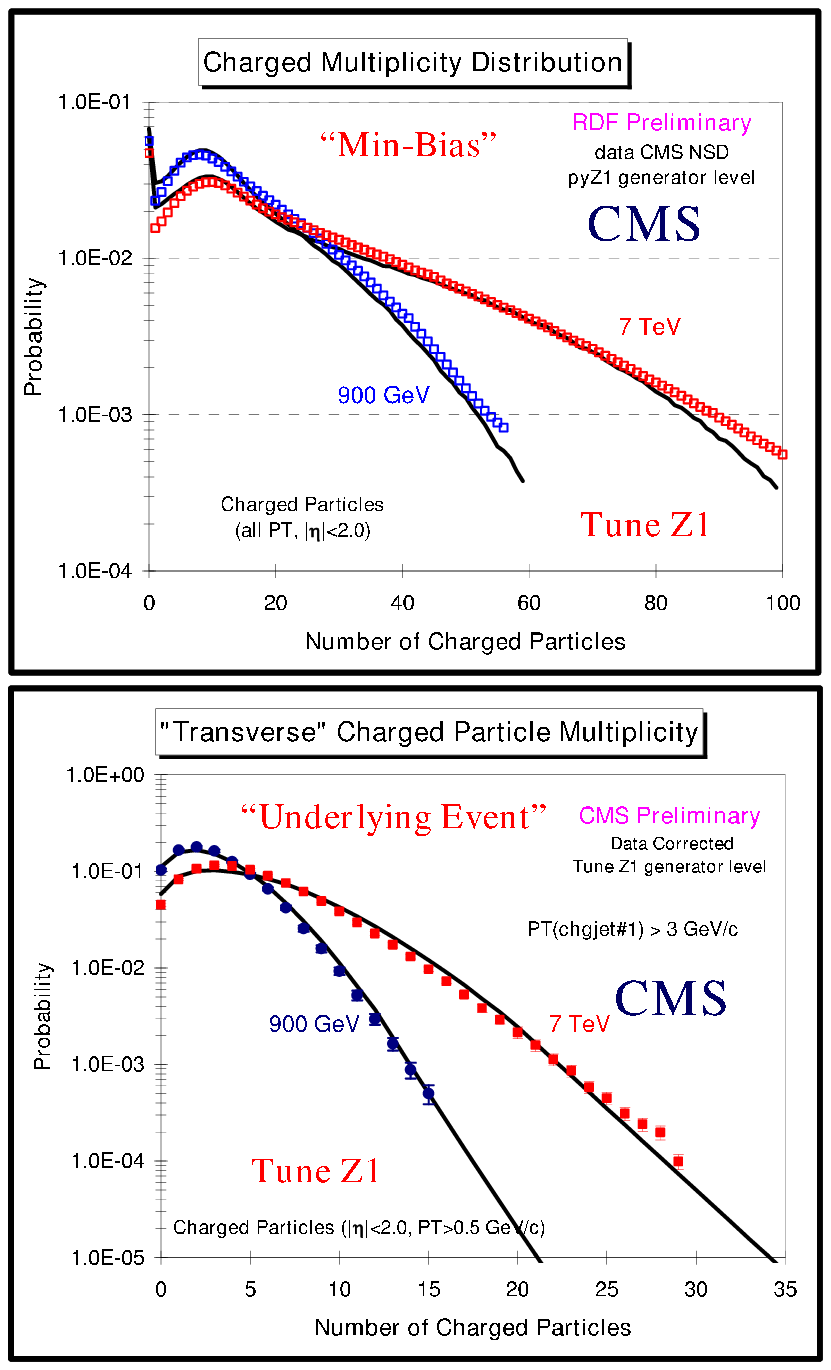}
\caption{\footnotesize
(\textit{top}) The non-single diffraction (NSD) data from CMS at $900\gev$ and $7\tev$ \cite{cmsmb2} on the charged particle multiplicity 
distribution (\etacut, all \pt) compared with PYTHIA Tune Z1. (\textit{bottom}) Data from CMS at $900\gev$ and $7\tev$ \cite{cmsue3} on the 
charged particle multiplicity distribution (\etacut, \ptlcut) in the \TR\ region as defined by the leading charged particle 
jet, \CJone, for \ptcj $> 3.0\gevc$ compared with PYTHIA Tune Z1.  The data have been corrected to the particle level and 
compared with Tune Z1 at the generator level.
}
\end{center}
\label{Zakopane_fig25}
\end{figure}

\section{PYTHIA Tunes and the LHC MB Data}

Since PYTHIA regulates both the primary hard scattering and the MPI with the same cut-off, $p_{T0}$, with PYTHIA one can model 
the overall "non-diffractive" (ND) cross section by simply letting the transverse momentum of the primary hard scattering go to 
zero.  The non-diffractive cross section then consists of BBR plus "soft" MPI with one of the MPI occasionally being hard. In this 
simple approach the UE in a hard-scattering process is related to MB collisions, but they are not the same.  Of course, to model 
MB collisions one must also add a model of single (SD) and double diffraction (DD).  This makes the modeling of MB much more 
complicated than the modeling of the UE. One cannot trust the PYTHIA 6.2 modeling of SD and DD.

Figure~21 shows the inelastic (INEL) data from ALICE and UA5 at $900\gev$ \cite{alice1} on the charged particle density, $dN/d\eta$ (all \pt) 
compared with PYTHIA Tune DW and Tune S320 \cite{skands1}.  Both these tunes are about $11\%$ below the data.  The INEL cross section is the sum 
of ND + SD + DD.  Fig.~22 shows the non-single diffraction (NSD) data from CMS $7\tev$ \cite{cmsmb1} on the charged particle density, $dN/d\eta$ 
(all \pt) compared with PYTHIA Tune DW.  The solid curve is NSD and the dashed curve is inelastic non-diffraction (ND) component. 
The NSD cross section is the sum of ND + DD.  Fig.~22 also shows the INEL data from ALICE at 900 GeV \cite{alice2} on the charged particle 
density, $dN/d\eta$, with $p_T>$ PTcut and with at lease one charged particle with $p_T>$ PTcut and $|\eta|<0.8$ for 
PTcut $=0.15\gevc$, $0.5\gevc$, and $1.0\gevc$ compared with PYTHIA Tune DW.  Tune DW was tuned to fit the Tevatron data 
with \ptlcut.  Two things change when we extrapolate from the Tevatron to the LHC.  Of course the center-of-mass energy changes, 
but also we have started looking at softer particles (\ie $p_T < 500\mevc$).  Fig.~22 shows that Tune DW does okay 
for \ptlcut, but does not produce enough soft particles below $500\mevc$.  One can also see that, at least in PYTHIA $6.2$, the 
modeling of SD and DD is more important at the lower \pt\ values.  

Figure~23 compares the CMS and ALICE charged particle densities, $dN/d\eta$, with PYTHIA 6.4 Tune Z1.  Tune Z1 does a better job at 
fitting the MB data than does Tune DW and it produced more soft particles below $500\mevc$ than does Tune DW.  However, Tune Z1 
does not fit the MB data perfectly.

Figure~24 compares the activity in the UE of a hard scattering process with an average MB collision.   The activity in the UE of a 
hard scattering process at $7\tev$ is roughly a factor of two greater than it is for an average MB collision and Tune Z1 describes 
this difference fairly well.  In PYTHIA this difference comes from the fact that there are more MPI in a hard scattering process 
than in a typical MB collision.  By demanding a hard scattering you force the collision to be more central (\ie smaller impact 
parameter), which increases the chance of MPI.

Figure~25 shows the data from CMS at $900\gev$ and $7\tev$ \cite{cmsmb2} on the charged particle multiplicity distribution (\etacut, all \pt) 
compared with PYTHIA Tune Z1 and the data from CMS at $900\gev$ and $7\tev$ on the charged particle multiplicity  
distribution in the \TR\ region as defined by the leading charged particle jet, \CJone, for \ptcj $>3.0\gevc$ compared with 
PYTHIA Tune Z1.  You are asking a lot of the QCD Monte-Carlo model when you expect it to simultaneously describe both MB and 
the UE in a hard scattering process.  I think it is amazing that Tune Z1 does as well as it does in describing both!

\section{Summary and Conclusions}

The PYTHIA $6.2$ Tune DW which was created from CDF UE studies at the Tevatron did a fairly good job in predicting the LHC UE 
data $900\gev$ and $7\tev$.  The behavior of the UE at the LHC is roughly what we expected.  Remember this is ``soft" QCD!  The 
LHC PYTHIA $6.4$ Tune Z1 does a very nice job describing the UE data at $900\gev$ and $7\tev$.  The UE is part of a hard 
scattering process.  MB collisions quite often contain no hard scattering and are therefore more difficult to model.  Since 
PYTHIA regulates both the primary hard scattering and the MPI with the same cut-off, $p_{T0}$, with PYTHIA one can model the 
overall non-diffractive (ND) cross section by simply letting the transverse momentum of the primary hard scattering go 
to zero.  In this approach the UE in a hard-scattering process is related to MB collisions, but they are not the 
same. Of course, to model MB collisions one must also add a model of single (SD) and double diffraction (DD).  Tune Z1 
does a fairly good job of simultaneously describing both MB and the UE in a hard scattering process.   I think it is amazing 
that it does as well on MB as it does!

There are a lot of factors of two floating around.  The charged particle density in the \TR\ region increases by about a factor 
of two in going from $900\gev$ and $7\tev$ (Fig.~15).  At $7\tev$ the charged particle density in the \TR\ region increases by 
about a factor of two in going from $p_T>500\mevc$ to $p_T>100\mevc$ (Fig. 18). The charged particle density in the \TR\ region 
is about a factor of two larger than the density of particles in a typical MB collision (Fig. 24).  All of these factors of two 
are described fairly well by PYTHIA Tune Z1.  PYTHIA $8$ \cite{pythia8} also does a fairly good job on many of the MB observables, 
but so far it does not fit the LHC UE data as well as Tune Z1.

In order to describe the bulk of the LHC MB data one must include a model of diffraction.  Experimentally, it is not possible 
to uniquely separate diffractive from non-diffractive collisions.  However, one can construct samples of ``diffraction enhanced MB" 
and ``diffraction suppressed MB" events and compare with the models.  The ``diffraction enhanced MB" samples are selected by 
requiring some type of rapidity gap \cite{cmsfwd,atlas3}.  We have learned that PYTHIA $6$ does a poor job of modeling of diffraction.  
PHOJET \cite{phojet} and PYTHIA $8$ do a better job with diffraction.    

In a very short time the experiments at the LHC have collected a large amount of data that can be used to study MB collisions 
and the UE in great detail.  This data can be compared with the Tevatron MB and UE data to 
further constrain and improve the QCD Monte-Carlo models we use to simulate hadron-hadron collision.  At present none of 
the tunes describe perfectly the UE data at both the Tevatron and the LHC.  However, I believe the tunes will continue to 
improve. We are just getting started!  The future will include more comparisons with PYTHIA $8$, HERWIG++ \cite{hw++}, and SHERPA \cite{sherpa}.


\end{document}